\documentstyle[emulateapj,onecolfloat,psfig]{article}

\long\def\comment#1{}

\def\la{\hbox{ \raise.35ex\rlap{$<$}\lower.6ex\hbox{$\sim$}\ }}
\def\ga{\hbox{ \raise.35ex\rlap{$>$}\lower.6ex\hbox{$\sim$}\ }}

\def\W2{{\cal W}}

\newcommand{\wjm}{\left(
			   \begin{array}{ccc}
	 \ell_1 & \ell_2  & \ell_3  \\
         m_1 & m_2  & m_3
		           \end{array}
                   \right)}

\newcommand{\wjma}[6]{\left(
			   \begin{array}{ccc}
	 #1 & #2  & #3  \\
         #4 & #5  & #6
		           \end{array}
                   \right)}

\newcommand{\xig}[2]{\xi({\bf e}_{#1}\!\cdot\! {\bf e}_{#2})}
\newcommand{\xigi}[2]{\xi^{-1}({\bf e}_{#1}\!\cdot\! {\bf e}_{#2})}

\newcommand{\dO}[1]{{\rm d}\Omega_{#1}}

\newcommand{\lamudeux}[5]{\lambda^{#5}_{ \left\{   \!\!\!  
                                       \tiny\begin{array}{cc}
         #1 \! & \! #2    \\
         #3 \! & \! #4 
		           \end{array} \!\!\! \right\}  }}

\newcommand{\lamu}[7]{\lambda^{#7}_{      \left\{  \! 
                                       \tiny\begin{array}{ccc}
         #1 \! & \! #2 \!  & \! #3 \\
         #4 \! & \! #5 \!  & \! #6
		           \end{array} \! \right\}   }}

\newcommand{\coefd}[5]{{\rm d}^{#5}_{ \left\{\!\! \tiny\begin{array}{ccc}
         \! #1 \!\! & \!\! #2 \! \\
         \! #3 \!\! & \!\! #4 \!
		           \end{array} \!\!\right\} }}
\newcommand{\coefdd}[9]{{\rm d}^{\tiny\begin{array}{ccc}
                        \! #7 \! & \! #8 \! & #9
                        \end{array}}_{ \left\{\!\! \tiny\begin{array}{ccc}
         \! #1 \!\! & \!\! #2 \!\!  & \!\! #3 \! \\
         \! #4 \!\! & \!\! #5 \!\!  & \!\! #6 \!
		           \end{array} \!\!\right\} }}

\def\la{\bigl\langle} \def\ra{\bigr\rangle}

\def\go{{\bf e}_1} \def\gt{{\bf e}_2}  \def\gth{{\bf e}_3}
\def\gf{{\bf e}_4} \def\gfi{{\bf e}_5} \def\gs{{\bf e}_6}
\def\gi{{\bf e}_i} \def\gj{{\bf e}_j}  \def\gk{{\bf e}_k}

\def\cd{\!\cdot\!}

\newcommand{\Ylm}[1]{Y_{\ell_{#1}}^{m_{#1}}}
\newcommand{\Ylmn}{Y_{\ell}^{m}}
\newcommand{\alm}[1]{a_{\ell_{#1}}^{m_{#1}}}
\newcommand{\almn}{a_{\ell}^{m}}

\def\be{\begin{equation}}
\def\bea{\begin{eqnarray}}
\def\ee{\end{equation}}  
\def\eea{\end{eqnarray}}

\begin{document}
\title{Best Unbiased Estimators for the Three-Point Correlators of the 
       Cosmic Microwave Background Radiation}
\author{Alejandro Gangui$^{1,2,3}$ and J\'er\^ome Martin$^{3}$}
\affil{$^{1}$Instituto de Astronom\'{\i}a y F\'{\i}sica del Espacio, 
Ciudad Universitaria, 1428 Buenos Aires, Argentina}
\affil{$^{2}$Dept. de F\'{\i}sica, Universidad de Buenos Aires, 
Ciudad Universitaria -- Pab. 1, 1428 Buenos Aires, Argentina}
\affil{$^{3}$DARC, Observatoire de Paris--Meudon, 
UMR 8629 CNRS, 92195 Meudon Cedex, France.}
\affil{E-mail: gangui@irup\'e.obspm.fr, martin@edelweiss.obspm.fr}
\begin{abstract}

Measuring the three-point correlators of the Cosmic Microwave
Background (CMB) anisotropies could help to get a handle on the level
of non-Gaussianity present in the observational datasets and therefore
would strongly constrain models of the early Universe.  However,
typically, the expected non-Gaussian signal is very small.  Therefore,
one has to face the problem of extracting it from the noise, in
particular from the `cosmic variance' noise.  For this purpose, one
has to construct the best unbiased estimators for the three-point
correlators that are needed for concrete detections of non-Gaussian
features.  In this article, we study this problem for both the CMB
third moment and the CMB angular bispectrum.  We emphasize that the
knowledge of the best estimator for the former does not permit one to
infer the best estimator for the latter and vice versa.  We present
the corresponding best unbiased estimators in both cases and compute
their corresponding cosmic variances.

\end{abstract}



\section{Introduction}
\footnote{A.G. would like to dedicate this article to the
       memory of his Ph.D. supervisor, Dennis William Sciama.}


\vspace{-0.2in}
\hspace{-0.07in}
The Cosmic Microwave Background (CMB) has been recognized as one of
the best tools for studying the early Universe (e.g. \cite{scott99}).
In particular, the statistical properties of the CMB anisotropies are
a powerful means to discriminate amongst the possible scenarios. This
is because, in general, different models predict different statistical
properties.  For example, the simplest models of inflation predict
that the temperature anisotropies should obey a Gaussian statistics
and therefore any non-vanishing measurement of a three-point
correlator (in a sense to be precised below) would automatically ruled
out such models, a very interesting result indeed.

{}From a practical point of view, measuring any non-Gaussianity in the
data is a very difficult task since the signal is typically very
small. Of course, this signal should be compared to the noise and what
really matters is the signal to noise ratio. The noise can have
many different origins including instrumental errors, foregrounds
contamination or incomplete sky coverage. Another source of error
is the so-called `cosmic variance'. Roughly speaking, it comes from
the fact that we only have access to one realization of the
temperature anisotropies whereas theoretical predictions are expressed
through ensemble averages. In a Gaussian model, for example, the mean
value of any three-point correlator has to vanish but this does not
guarantee that a concrete detection of a non-zero signal on the sky
would be in contradiction with the model (\cite{sc91,sr93}).  The
important point is that the cosmic variance can dominate the other
sources of error, as this is in fact the case for the two-point
correlators on large angular scales. Therefore, if one wants to
unveil non-Gaussianity, it is necessary to address the cosmic variance
problem for the three-point correlators.
The usual way to deal with this problem is to construct estimators by
performing spatial averages on the celestial sphere and to find the
one which has the smallest possible variance. The aim of this paper is
then to find the best unbiased estimators both for the third moment
$\bigl\langle a_{\ell_1 }^{m_1} a_{\ell_2}^{ m_2} a_{\ell_3}^{ m_3}
\bigr\rangle$ and for the angular bispectrum ${\cal C}_{\ell_1 \ell_2
\ell_3 }$ and to display the corresponding cosmic variances.

Recently, there has been a lot of activity in the subject triggered by
the finding that non-Gaussianities are present in the 4-yr COBE-DMR
data (\cite{Feretal98,Pando98}).  Further analyses have confirmed this
result (e.g. \cite{BroTeg99}). However, soon after, it was
demonstrated by \cite{BZG} (1999) that the non-Gaussian signal is
driven by the 53 GHz data. This systematic artifact in the CMB 
maps rejects
a possible cosmological origin.  More generally, it is clear that the
presence of foregrounds (\cite{fo1,fo2}) renders difficult the
detection of a genuine non-Gaussian signal.  Nevertheless, one should
expect non-Gaussian features to be present in the CMB anisotropy
datasets. These could be produced in the early Universe during
inflation either because the initial conditions are non-Gaussian
themselves [i.e. the quantum initial state is not the vacuum
(\cite{MRS,CBM})] or owing to the existence of couplings between
different perturbation modes at the non-linear level
(\cite{Ganetal94,ga94,LM}). In the context of slow-roll inflation, the
CMB bispectrum has recently been studied in
(\cite{GanMar00,WanKam99}).  Even if non-Gaussianities are not
primordial in origin, they will nevertheless arise during later stages of
evolution.  In this context, the Rees-Sciama effect will build up a
small but non-vanishing signal (\cite{LuoSch93,molle,mun}).  Also,
cosmic topological defects of the vacuum, like strings and textures,
are amongst the best motivated sources for non-Gaussian features
(\cite{bouchet,fema97,avel,ga95,gamo96}).  Regarding secondary
sources, Goldberg \& Spergel (1999) and Spergel \& Goldberg (1999)
have recently calculated the angular bispectrum due to second order
gravitational effects like the correlation of lensing of CMB photons
and secondary anisotropies coming from the Integrated Sachs Wolfe
effect and thermal Sunyaev-Zel'dovich effect. In the same line, Cooray
and Hu (1999) have taken into account further additional
contributions to the bispectrum in the presence of reionization.
Other approaches to the study of non-Gaussian
features include preferred-direction statistics for sky maps
(\cite{bs99}), the three-point correlation function
(\cite{Falk93,hinshaw94,Ganetal94,hinshaw95}), 
lensing statistics (\cite{bernardo,wini,zal}), 
the genus and Euler-Poincar\'e statistics
(\cite{coles89,gott90,smoot94}), peak statistics
(\cite{be87,ko95,ko96}), correlation function of peaks (\cite{hs99}),
Minkowski functionals (\cite{wk98}) and wavelet analyses
(\cite{popa98,hobson98}).

This article is organized as follows. In the next section, the general
strategy for finding best estimators is exposed. As a warm up, in the
third section, we implement this strategy for the two-point
correlators. The fourth section is the core of the article. 
There we explicitly derive, for the first time, the best
unbiased estimator for the angular bispectrum and show its
corresponding variance. Except for an overall normalization factor,
this estimator turns out to be the one already employed 
by Ferreira et al. (1998) and other authors recently. Our result 
places their choice on a firm basis.  
Next, we find the expression for the best
unbiased estimator for the third moment.  An earlier study was
performed in (\cite{Hea98}); however, our findings go beyond the
results obtained in that article and, moreover, are explicit.  
Moreover, we present its corresponding variance.
In addition, we also emphasize that the knowledge of the best estimator
for the third moment does not allow one to infer the best estimator
for the angular bispectrum and vice versa.  In the last section, we
briefly present our main conclusions.  We finish up with a short
Appendix which includes formulae related to the inverse two-point
correlation function.

\section{General strategy for finding the best estimator}

In this section, we expose the cosmic variance problem from the
viewpoint of the theory of cosmological perturbations of
quantum-mechanical origin and describe the method of the best unbiased
estimators. This theory rests on the principles of general relativity
and quantum field theory. At the beginning of the inflationary phase
(\cite{G,Lin1,AS,Lin2}) the Friedmann-Lema{\^\i}tre-Robertson-Walker
background spacetime already behaves classically whereas the
excitations of the metric around this background are still quantum
mechanical in nature. Technically, this means that the perturbed
metric must be considered as a quantum operator.  This operator either
represents density perturbations or gravitational waves. In each case,
the quantization can be carried out in a consistent way
(\cite{MC,Haw,Staro,BST,MFB92,Grigw,MS1}, 1999).  
Then, the (zero-point)
quantum fluctuations, which are the seeds of the cosmological
perturbations, are amplified during inflation owing to the
particle-creation phenomenon or squeezing effect (\cite{GriSid}).
Next, these primordial fluctuations give rise to the large scale
structures and to the CMB anisotropies observed today in our Universe.

One should also discuss the choice of the quantum state in which the
metric operator is placed. Obviously, it is not possible to prepare
the initial state of the Universe and therefore the choice of the
quantum state of the perturbations is {\em a priori} free unless some
theory of the initial conditions is provided [for example, quantum
cosmology (\cite{Halli})]. Usually, it is assumed that the initial
state is the vacuum although different hypothesis are possible
(\cite{BH,MRS,CBM}). If the initial state is the vacuum, then the
corresponding statistical properties are Gaussian. This is because the
ground-state wave function of an harmonic oscillator is a Gaussian. It
is possible to avoid this general conclusion either by considering
non-linear cosmological perturbations or by assuming that the initial
state is a non-vacuum state. We have recently investigated the first
possibility in (\cite{GanMar00}).  The second possibility has been
studied by Martin et al. (1999). In the latter case, non-Gaussianity is
likely to be significant only for relatively small angular scales.

It should be emphasized that the mechanism described previously is
deeply rooted in the quantum-mechanical nature of the gravitational
field. The observable quantities calculated in this framework are
always proportional to the Planck length.  In other words, if
observations confirm the full set of inflationary predictions then the
fact that ${\rm \delta }T/T \neq 0$ would be a direct observational
consequence of quantum gravity. 

The quantum-mechanical origin of the anisotropies in the framework of
inflation raises also profound problems of interpretation. One should
not think that these problems are purely theoretical. On the contrary,
they have consequences with regards to the experimental strategy that
one should follow in order to extract as much informations as possible
from the data. The fluctuations in the CMB effective 
temperature are linked to
the perturbed metric as shown for the first time by Sachs and Wolfe.
Therefore, the fact that the perturbed metric is an operator implies
that the primordial fluctuations in the temperature must also be
considered as a quantum operator.  The observables are often expressed
as $n$-point correlation functions of the operator $\hat{\Delta }({\bf
e}) \equiv \hat{ \delta T / T} ({\bf e})$ in the arbitrary state 
$\vert\Psi\rangle$
\begin{equation}
\label{ncorr}
\xi _n({\bf e}_1, \cdots ,{\bf e}_n)\equiv 
\langle \Psi \vert \hat{\Delta }({\bf e}_1) \cdots 
\hat{\Delta }({\bf e}_n) \vert \Psi \rangle ,
\end{equation}
where ${\bf e}_i$'s are arbitrary directions on the celestial sphere.
In the following, we will also use the notation $\xi ({\bf
e}_1\!\cdot\!{\bf e}_2) \equiv \xi _2({\bf e}_1,{\bf e}_2)$.
According to the postulates of quantum mechanics, the previous
theoretical predictions should be confronted to experiment in the
following way. The same experiment should be performed $N$ times
giving each time different outcomes $q_i$. If the quantity $(1/N)\sum
_{i=1}^Nq_i$ goes to the corresponding quantum expectation value when
$N$ goes to infinity, the theoretical prediction is said to be
`compatible with experiment'. This is the core of the problem in
cosmology: we only have access to one realization, i.e. one map of the
CMB sky and that means that $N$ is fixed and equal to one. Therefore,
the question arises as to how we can verify the theoretical
predictions of the theory of quantum-mechanical cosmological
perturbations. This is a way of stating the cosmic variance problem. 
It is a
fundamental limitation in the sense that it remains even when other
limitations like instrumental errors or low angular resolution have
been fully mastered.

The usual method to deal with this problem is to replace quantum
averages with spatial averages over the celestial sphere. 
Suppose we wish to measure
$\xi _n({\bf e}_1, \cdots {\bf e}_n)$. 
(Of course, the discussion could
also be applied to quantities other than correlation functions). 
The first
step is to introduce a new operator, the estimator $\hat{\cal
E}(\xi _n)$ of $\xi _n$, defined as
\begin{equation}
\label{defE_K}
\hat{\cal E}(\xi _n)\equiv \int \cdots \int {\rm d}\Omega _1 \cdots 
{\rm d}\Omega _n E(\xi _n)({\bf e}_1\cdots {\bf e}_n)  
\hat{\Delta }({\bf e}_1)\cdots \hat{\Delta }({\bf e}_n),
\end{equation}
where $E(\xi _n)({\bf e}_1\cdots {\bf e}_n)$ is a weight function to
be determined. Clearly, $\hat{\cal E}(\xi _n)$ is
defined through a spatial average. The second step is to require that
the estimator is unbiased, i.e.
\begin{equation}
\label{meanE}
\langle \Psi \vert \hat{\cal E}(\xi _n)\vert \Psi \rangle = \xi _n.
\end{equation}
In general, this restricts the class of functions $E(\xi _n)$ 
allowed. The fact that the mean value of the estimator 
be equal to the quantity we are seeking does not guarantee that 
each outcome will be for sure $\xi _n$. The third and 
final step is then to find the function $E(\xi _n)$ such that the 
variance (squared) of $\hat{\cal E}(\xi _n)$, i.e. 
\be
\label{vari0}
\sigma^2_{\hat{\cal E}(\xi _n)} =
\la \hat{\cal E}(\xi _n)\hat{\cal E}(\xi _n) \ra -
\la \hat{\cal E}(\xi _n) \ra^2
\ee
be as small as possible, taking into account the constraint given by
Eq. (\ref{meanE}). The corresponding estimator is then called the {\it best
unbiased estimator}. Mathematically, this requirement is expressed
through the following variation equation
\begin{equation}
\label{minicons}
{\rm \delta }\biggl[\sigma _{\hat{\cal E}(\xi _n)}^2-\lambda
\biggl(\langle \Psi \vert \hat{\cal E}(\xi _n)\vert \Psi \rangle -\xi
_n \biggr)\biggr]=0,
\end{equation}
where one has introduced a Lagrange multiplier $\lambda$ which can
then be determined from the previous equation and the constraint
itself. Once we have $\lambda$, we plug it into Eq. (\ref{minicons})
and this completely fixes $E(\xi _n)$ and hence the corresponding best
estimator.  In turn, its variance can now be calculated. If this one
vanishes then we are sure that each outcome is $\xi _n$ and
from one realization we can determine the $n$-point correlation
function. In this case, $\hat\Delta({\bf e})$ 
is said to be ergodic, i.e. 
ensemble or quantum averages coincide with spatial
averages. Unfortunately, one can show that this cannot be the case on
the two-dimensional sphere (\cite{GriM}). Otherwise, we have found the
weight function $E(\xi _n)$ which leads to the smallest non-vanishing
variance. If the variance is small enough, each outcome will be
concentrated around the mean value and with just one realization we
have good chances to get a reasonable estimate of the correlation
function. The typical error made in considering that one given outcome
is equal to the mean value is characterized by the variance of the
estimator, see Fig. 1.
\begin{figure}[t]
\centerline{\psfig{file=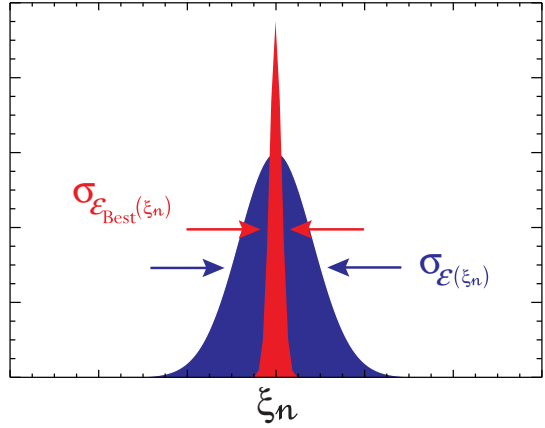,width=3.6in}}
\caption{ Sketch comparing the variances of an arbitrary estimator of
$\xi_n$ with the best estimator of the same quantity.  The widest
distribution does not permit an accurate determination of $\xi_n$ due
to its large variance whilst the narrow distribution corresponds to
the best (unbiased) estimator and possesses the smallest possible
variance. As a consequence, one given realization (i.e. our CMB sky)
will most probably be closest to the mean value in the second case
rather than in the first one.}
\label{fig0}
\end{figure}

All the above analysis performed for the best (quantum) estimators can
equally well be reproduced in the case where the anisotropies are due
to an underlying stochastic process although the former is generally
physically best motivated. In that case, quantum averages $\langle
\Psi \vert \cdots \vert \Psi \rangle$ are just replaced with
stochastic averages $\langle \cdots \rangle$.  In the following, we
will drop out the `hat' symbol and consider that the different
quantities are either operators or stochastic processes. In the same
manner, we will denote an ensemble average by the symbol $\langle
\cdots \rangle$ having in mind that this means either quantum or
classical averages.

Let us now describe the relevant quantities to estimate. It is
convenient to expand the temperature fluctuations over the basis of
spherical harmonics according to
\begin{equation}
\label{expSW}
\Delta ({\bf e})=\sum _{\ell m}
a_{\ell  }^{m}\Ylm{}({\bf e}).
\end{equation}
This equation assumes a complete sky coverage. Implementation of the 
method for the incomplete (galaxy-cut) sky can be performed by using 
the basis introduced in (\cite{Gor94}). Once a specific model is 
given, the statistical properties of the
$a_{\ell }^{m}$'s are determined. Since $\Delta ({\bf e})$ is real,
the $a_{\ell }^{m}$'s must satisfy $a_{\ell }^{m*}= (-1)^ma_{\ell
}^{-m}$. Without restricting the generality of the underlying physics,
the first three moments can be written as
\begin{equation}
\label{propa}
\bigl\langle a_{\ell  }^{m} \bigr\rangle =0, \quad 
\bigl\langle a_{\ell _1 }^{m_1}a_{\ell _2 }^{m_2*} \bigr\rangle =
{\cal C}_{\ell_1}\delta _{\ell_1\ell_2}\delta _{m_1m_2}, \quad 
\bigl\langle a_{\ell_1 }^{m_1} a_{\ell_2}^{ m_2} a_{\ell_3}^{ m_3}
\bigr\rangle = 
\left(^{\ell_1~\,\;\ell_2~\,\;\ell_3}_{m_1~m_2~m_3}\right)
{\cal C}_{\ell_1 \ell_2 \ell_3 },
\end{equation}
where $\left(^{\ell_1~\,\;\ell_2~\,\;\ell_3}_{m_1~m_2~m_3}\right)$ is
a Wigner 3$j$-symbol. The second equation ensures the isotropy of the
CMB.  The quantity $\bigl\langle a_{\ell _1 }^{m_1}a_{\ell _2 }^{m_2*}
\bigr\rangle$ is the second moment of the $\almn$'s and ${\cal
C}_\ell$ is usually called the angular spectrum.  In the third
equation, the quantity $\bigl\langle a_{\ell_1 }^{m_1} a_{\ell_2}^{
m_2} a_{\ell_3}^{ m_3} \bigr\rangle$ is the third moment while 
${\cal C}_{\ell_1 \ell_2 \ell_3 }$ is called the angular
bispectrum.  For $\ell_1= \ell_2= \ell_3 =\ell$, this quantity is
generally written as ${\cal B}_{\ell }\equiv {\cal C}_{\ell \ell \ell
}$.  The presence of the Wigner 3$j$-symbol guarantees that the third
moment vanishes unless $m_1+m_2+m_3=0$ and $|\ell _i-\ell_j| \le \ell
_k \le \ell _i+\ell _j$.  Moreover, invariance under spatial
inversions of $\xi_3$ implies an additional `selection rule'
(\cite{luo94,GanMar00}), 
$\ell _1 +\ell _2 +\ell _3=\mbox{even}$, in order for the third moment
not to vanish.  Finally, from this last relation and using standard
properties of the 3$j$-symbols, it follows that the angular bispectrum
is left unchanged under any arbitrary permutation of the indices
$\ell_i$.  

We will need the higher moments as well.  Since departures
from Gaussianity are expected to be small (specially on large angular
scales), higher moments will be calculated in the mildly non-Gaussian
approximation.  Within this approximation we can write $a_{\ell}^{m} =
a_{\ell}^{m (0)} + \epsilon \, a_{\ell}^{m (1)} + {\cal O}(\epsilon^2)$
where $a_{\ell}^{m (0)}$ is a Gaussian random variable and the
expansion parameter $\epsilon$ is small.  In the following, each
moment will be calculated to the first non-vanishing order in
$\epsilon$.  For example, the fourth moment yields $\bigl\langle
a_{\ell_1 }^{m_1} a_{\ell_2}^{ m_2} a_{\ell_3}^{ m_3} a_{\ell_4}^{
m_4}\bigr\rangle = \bigl\langle a_{\ell_1 }^{m_1 (0)}
a_{\ell_2}^{ m_2 (0)} a_{\ell_3}^{ m_3 (0)} a_{\ell_4 }^{ m_4
(0)}\bigr\rangle + {\cal O}(\epsilon)$. As a consequence, the
connected fourth moment can be neglected because it is of higher order
than the Gaussian part.  The `$^{(0)}$' label will be dropped out
hereafter.  Therefore, in the mildly non-Gaussian approximation we can
write
\begin{eqnarray}
\label{4a}
\bigl\langle a_{\ell_1 }^{m_1} a_{\ell_2}^{ m_2} a_{\ell_3}^{ m_3*}
 a_{\ell_4}^{ m_4*}\bigr\rangle &\approx &
\bigl\langle a_{\ell _1 }^{m_1}a_{\ell _2 }^{m_2} \bigr\rangle 
\bigl\langle a_{\ell _3 }^{m_3*}a_{\ell _4 }^{m_4*} \bigr\rangle +
\bigl\langle a_{\ell _1 }^{m_1}a_{\ell _3 }^{m_3*} \bigr\rangle 
\bigl\langle a_{\ell _2 }^{m_2}a_{\ell _4 }^{m_4*} \bigr\rangle +
\bigl\langle a_{\ell _1 }^{m_1}a_{\ell _4 }^{m_4*} \bigr\rangle 
\bigl\langle a_{\ell _2 }^{m_2}a_{\ell _3 }^{m_3*} \bigr\rangle \nonumber \\
&=&
(-1)^{m_2+m_4}{\cal C}_{\ell_1}{\cal C}_{\ell_3}\delta _{\ell_1\ell_2}
\delta _{m_1,-m_2}\delta _{\ell_3\ell_4}\delta _{m_3,-m_4}+
{\cal C}_{\ell_1}{\cal C}_{\ell_2}
\delta _{\ell_1\ell_3}\delta _{m_1m_3}
\delta _{\ell_2\ell_4}\delta _{m_2m_4}
\nonumber \\
&+&
{\cal C}_{\ell_1}{\cal C}_{\ell_2}
\delta _{\ell_1\ell_4}\delta _{m_1m_4}
\delta _{\ell_2\ell_3}\delta _{m_2m_3} .
\end{eqnarray}
The fifth moment could be determined in a similar way but we will not
need this quantity in the following. Finally, the sixth moment can be
expressed as
\begin{equation}
\label{6a}
\bigl\langle a_{\ell_1 }^{m_1} a_{\ell_2}^{ m_2} a_{\ell_3}^{ m_3}
a_{\ell_4}^{ m_4*} a_{\ell_5}^{ m_5*}a_{\ell_6}^{ m_6*}\bigr\rangle \approx 
\bigl\langle a_{\ell_1 }^{m_1} a_{\ell_2}^{ m_2}\bigr\rangle 
\bigl\langle a_{\ell_3 }^{m_3} a_{\ell_4}^{ m_4*}\bigr\rangle 
\bigl\langle a_{\ell_5}^{m_5*} a_{\ell_6}^{ m_6*}\bigr\rangle 
+\mbox{ 14 additional permutations }. 
\end{equation}
Although the explicit expression is not particularly illuminating, 
the last equation will be useful for the calculation of 
the variance when dealing with the three-point correlators below.
In particular, one can write
\begin{eqnarray}
\label{Luo}
\bigl\langle a_{\ell_1 }^{m_1} a_{\ell_2}^{ m_2} a_{\ell_3}^{ m_3}
             a_{\ell_1}^{ m_1*} a_{\ell_2}^{ m_2*}a_{\ell_3}^{ m_3*}
\bigr\rangle
&=& 
{\cal C}_{\ell_1}{\cal C}_{\ell_2}{\cal C}_{\ell_3}+2{\cal C}_{\ell_1}^3
\delta _{\ell_1\ell_2\ell_3}
(\delta _{m_1m_3}+\delta _{m_1 -m_3})(\delta _{m_1m_2}+\delta _{m_1 -m_2})
\nonumber \\
&+&
{\cal C}_{\ell_1}{\cal C}_{\ell_2}^2
\delta _{\ell_2\ell_3}(\delta _{m_2m_3}+\delta _{m_2 -m_3})
+
{\cal C}_{\ell_2}{\cal C}_{\ell_3}^2
\delta _{\ell_3\ell_1}(\delta _{m_1m_3}+\delta _{m_1 -m_3})
\nonumber \\
&+&
{\cal C}_{\ell_3}{\cal C}_{\ell_1}^2
\delta _{\ell_1\ell_2}(\delta _{m_1m_2}+\delta _{m_1 -m_2}),
\end{eqnarray}
where the symbol $\delta _{\ell_1\ell_2\ell_3}$ vanishes 
unless $\ell_1=\ell_2=\ell_3$ in which case it is one. 
This equation coincides with 
Eq. (24) of (\cite{luo94}) 
provided the undefined symbol $\delta _{m_1m_2m_3,0}$ 
written in that work has the meaning 
$\delta _{m_1m_2m_3,0}\equiv (\delta _{m_1m_3}
+\delta _{m_1 -m_3})(\delta _{m_1m_2}+\delta _{m_1 -m_2})$. 

As we mentioned in the Introduction, in this article we are mainly
interested in finding the best unbiased estimators for the two
following quantities: the third moment $\bigl\langle a_{\ell_1 }^{m_1}
a_{\ell_2}^{ m_2} a_{\ell_3}^{ m_3}\bigr\rangle $ and the angular
bispectrum ${\cal C}_{\ell_1 \ell_2 \ell_3 }$.  These 
are related to the three-point correlation 
function $\xi _3$. It is clear that, as mentioned above, the very 
same method could also be utilized for the computation of $\xi _n$ 
with $n$ arbitrary. Before addressing this question, however, we 
will first treat the analogous
quantities related to the two-point correlation function, namely
$\bigl\langle a_{\ell_1 }^{m_1} a_{\ell_2}^{ m_2*} \bigr\rangle $ and
${\cal C}_{\ell }$, the main purpose being to illustrate concretely the
tactics presented above in a case where everything can be calculated
easily.  This will be used as a guideline for the case of the
three-point correlators.
 
\section{Two-point correlators}

\subsection{Best estimator for the angular spectrum ${\cal C}_{\ell}$}

The definition of the estimator ${\cal E}({\cal C}_{\ell })$ is given by 
our general prescription, see Eq. (\ref{defE_K})
\begin{equation}
\label{estiC_l}
{\cal E}({\cal C}_{\ell})\equiv \int \int {\rm d}\Omega _1 {\rm d}\Omega _2 
E^{\ell}({\bf e}_1,{\bf e}_2) \Delta ({\bf e}_1)\Delta({\bf e}_2),
\end{equation}
where $E^{\ell}({\bf e}_1,{\bf e}_2)$ is the weight function. The angular 
spectrum ${\cal C}_{\ell }$ is a real quantity and its
estimator ${\cal E}({\cal C}_{\ell})$ must also be real. 
Therefore, the weight function can be taken real.
{}From the previous definition, it is clear that the antisymmetric part of 
$E^{\ell}(\go,\gt)$ does not contribute to the estimator. Then, 
we can replace $E^{\ell}(\go,\gt)$ in ${\cal E}({\cal C}_{\ell})$
by its symmetrized expression $E_{\rm S}^{\ell}(\go,\gt) = (1/2)
[E^{\ell}(\go,\gt)+ E^{\ell}(\gt,\go)]$. 
At this stage, two methods can be applied. Either we
work directly with the weight function or we expand it over the
spherical harmonics basis and try to determine the coefficients of this 
expansion. Clearly, both paths are equivalent and can be followed for 
any estimator. Here we employ the second method, leaving the first one 
for the determination of the second-moment estimator 
considered in the next subsection. Therefore, we write the 
weight function as 
\begin{equation}
\label{expomC_l}
E^{\ell}_{\rm S}({\bf e}_1,{\bf e}_2)=\sum _{\ell_1 m_1}\sum _{\ell_2 m_2}
\coefd{\ell_1}{\ell_2}{m_1}{m_2}{\ \ell}Y_{\ell _1}^{m_1}({\bf e}_1)
Y_{\ell _2}^{m_2}({\bf e}_2).
\end{equation}
The reality and symmetry properties of the weight function 
$E_{\rm S}^{\ell}$ imply that the complex coefficient of the expansion must 
satisfy, respectively
\begin{equation}
\label{propd}
\coefd{\ell_1}{\ell_2}{m_1}{m_2}{\ \ell*}=
(-1)^{m_1+m_2}\coefd{\ell_1}{\ell_2}{-m_1}{-m_2}{\ \ell}
\quad 
,
\quad 
\coefd{\ell_1}{\ell_2}{m_1}{m_2}{\ \ell}=
\coefd{\ell_2}{\ell_1}{m_2}{m_1}{\ \ell}.
\end{equation}
Inserting the expression of the weight function (\ref{expomC_l}) into
the general definition of the estimator (\ref{estiC_l}) and using
standard properties of the spherical harmonics, one gets
\begin{equation}
\label{esti2pt}
{\cal E}({\cal C}_{\ell})=\sum _{\ell _1m_1}\sum _{\ell _2m_2}
\coefd{\ell_1}{\ell_2}{m_1}{m_2}{\ \ell *}a_{\ell _1}^{m_1}a_{\ell _2}^{m_2}.
\end{equation}
Our first move is now to require that $\langle {\cal E}({\cal C}_{\ell}) 
\rangle ={\cal C}_{\ell}$. 
Using the second of Eqns. (\ref{propa}), we find that the 
coefficients $d$ must fulfill the following constraints 
\begin{equation}
\label{consd}
\sum _{m_1}(-1)^{m_1}
\coefd{\ell_1}{\ell_1}{-m_1}{m_1}{\ \ell }
=\delta ^{\ell }_{\ell_1}.
\end{equation}
All the estimators satisfying this condition are unbiased. However, it
is clear that this does not completely determines the estimator but
just a class of estimators. 
Our second move is to calculate the variance. 
Using Eq. (\ref{4a}), a straightforward computation gives
\begin{equation}
\label{varCl}
\sigma ^2_{{\cal E}({\cal C}_{\ell })}=
2\sum _{\ell _1m_1}\sum _{\ell _2m_2}
\coefd{\ell_1}{\ell_2}{m_1}{m_2}{\ \ell *}
\coefd{\ell_1}{\ell_2}{m_1}{m_2}{\ \ell}
{\cal C}_{\ell _1}{\cal C}_{\ell _2}.
\end{equation}
This quantity is obviously positive. From this expression, one sees
that the imaginary part of the coefficients $d$ only increases the
variance. Since a vanishing $\Im({\rm d})$ satisfies the constraint equation, 
we can consider that the $d$'s are real. 

Our third move is to
minimize the variance taking into account the constraint. For this
purpose we introduce a set of Lagrange multipliers $\lambda ^{\ell }_{\ell
_1}$ (i.e. one Lagrange multiplier per constraint since $\ell $ must
be seen as a fixed index) and require that
\begin{equation}
\label{mini}
\delta \biggl[\sigma ^2_{{\cal E}({\cal C}_{\ell })}
+\sum _{\ell_1}\lambda ^{\ell }_{\ell _1}
\biggl(\sum _{m_1}(-1)^{m_1}
\coefd{\ell_1}{\ell_1}{-m_1}{m_1}{\ \ell}
-\delta ^{\ell }_{\ell _1}\biggl)\biggr]=0 .
\end{equation}
The definition of the variation $\delta$ must respect the symmetry 
properties of the coefficients $d$; we take  
\begin{eqnarray}
\label{defvar2pts}
{\delta \coefd{\ \ell_1  \ }{ \ \ell_2  \ }{\ m_1  \ }{ \ m_2  \ }
              {\ \ell}\over
\delta  \coefd{\ \ell_1' \ }{ \ \ell_2' \ }{\ m_1' \ }{ \ m_2' \ }
              {\ \ell}}
\equiv 
\frac{1}{2}&\biggl\{ &\frac{1}{2}\delta _{\ell _1 \ell_1'}\delta _{\ell _2 \ell_2'}
\biggl[\delta _{m_1 m_1'}\delta _{m_2 m_2'}+(-1)^{m_1+m_2}
\delta _{m_1 -m_1'}\delta _{m_2 -m_2'}\biggr]\nonumber \\
&+& \frac{1}{2}\delta _{\ell _1 \ell_2'}\delta _{\ell _2 \ell_1'}
\biggl[\delta _{m_1 m_2'}\delta _{m_2 m_1'}+(-1)^{m_1+m_2}
\delta _{m_1 -m_2'}\delta _{m_2 -m_1'}\biggr]\biggr\}.
\end{eqnarray}
Although it is not compulsory to use this equation, since a naive
definition of the variation would lead to the same final result
(\cite{GriM}), it is nevertheless interesting to utilize it as a warm
up for what will be done for the angular bispectrum. The variation
leads to the following relation between the coefficients $d$ and the
Lagrange multipliers
\begin{equation}
\label{resumini}
4\coefd{\ell_1}{\ell_2}{m_1}{m_2}{\ \ell}{\cal C}_{\ell _1}{\cal C}_{\ell _2}
+(-1)^{m_1}\lambda ^{\ell }_{\ell _1}\delta _{\ell _1 \ell_2}\delta _{m_1 -m_2}
=0 .
\end{equation}
We see that the choice of the $\lambda^\ell$'s is not free; it is
fixed by the variation itself. Using the constraint in this equation,
we find $\lambda ^{\ell }_{\ell _1}=-4[{\cal C}_{\ell _1}^2/(2\ell
_1+1)] \delta ^{\ell }_{\ell _1}$. Having determined what the Lagrange
multipliers are, the problem is completely solved. It is now
sufficient to re-introduce this value for $\lambda ^{\ell }_{\ell _1}$
in Eq. (\ref{resumini}), get the coefficients $d$ and, from this, also
the best unbiased estimator: the weight function is then given by
$E_{\rm S , Best}^{\ell }({\bf e}_1,{\bf e}_2)=(1/4\pi ) P_{\ell }(
{\bf e}_1 \cdot {\bf e}_2)$ and the estimator itself by (see also
\cite{GriM})
\begin{equation}
\label{estmulti}
{\cal E}_{\rm Best}({\cal C}_{\ell})=\frac{1}{2\ell+1}
\sum _{m=-\ell}^{\ell}a_{\ell }^{m}a_{\ell }^{m*}.
\end{equation}
The variance of this estimator is the well-known `cosmic variance'
\begin{equation}
\label{cosvar2pt}
\sigma ^2_{{\cal E}_{\rm Best}({\cal C}_{\ell})}=
\frac{2{\cal C}_{\ell}^2}{2\ell +1} .
\end{equation}
One remark is in order here. The cosmic variance is usually obtained
in the following way: the previous estimator appears naturally from
Eqns. (\ref{propa}) and it is usually assumed that the $a_{\ell
}^{m}$'s are Gaussian random variables. In this case, the estimator
(\ref{estmulti}) has a $\chi ^2$ probability density function and from
this the cosmic variance can be easily recovered. The proof presented
above is by no means equivalent to this naive derivation. There are
many unbiased estimators and, {\em a priori}, nothing guarantees that
the simplest one is the best one, i.e. the naive derivation is {\em not}
sufficient to prove that the estimator (\ref{estmulti}) is the best
one. This can be proven only along the lines described above.
Moreover, we do not need to assume that the $a_{\ell }^{m}$'s obey a
Gaussian statistics. Only the mildly non-Gaussian assumption is
necessary for the calculation of the four-point correlators.

\subsection{Best estimator for the second moment ${\cal C}_{\alpha }$}

In this section, we discuss the best estimator for 
${\cal C}_{\alpha }\equiv \bigl\langle a_{\ell_1 }^{m_1} 
a_{\ell _2 }^{m_2*} \bigr\rangle$. 
Greek letters will always be employed for collective-index notation, like 
$\alpha\equiv
{   \left\{   \!\!  \tiny \begin{array}{cc} 
\! \ell_1 \!\! & \!\! \ell_2 \! \\  
\! m_1    \!\! & \!\! m_2    \! 
\end{array} \!\! \right\}  }$ 
or 
$\alpha '\equiv
{   \left\{   \!\!  \tiny \begin{array}{cc}
\! \ell_1' \!\! & \!\! \ell_2' \! \\
\! m_1'    \!\! & \!\! m_2'    \!
\end{array} \!\! \right\}  }$. 
Unlike in the foregoing subsection, we find the weight
function directly without expanding it over the spherical harmonics
basis.  This method is closer to the one used by Heavens (1998).
We also show that the best estimator of ${\cal C}_{\alpha }$
cannot be deduced from the best estimator of the angular spectrum
${\cal C}_\ell$ obtained in the previous subsection as one could
naively think.

Let us start with a couple of definitions. First, the quantity 
\begin{equation}
\label{defR}
R^{\alpha }({\bf e}_1,{\bf e}_2)\equiv Y_{\ell _1}^{m_1}({\bf e}_1)
Y_{\ell _2}^{m_2*}({\bf e}_2) ,
\end{equation}
which is complex and non-symmetric with regards to the position of the
complex conjugate symbol `$^*$'. 
This last property comes from the definition of
$\bigl\langle a_{\ell _1 }^{m_1}a_{\ell _2 }^{m_2*} \bigr\rangle $
itself and it would be the case for any even-point correlator. Then,
regarding complex conjugation, there is a slight difference between
the even- and the odd-point correlators. The real part of $R^{\alpha
}({\bf e}_1,{\bf e}_2)$ will be noted
\begin{equation}
\label{defRR}
R^{\alpha }_{\rm R}({\bf e}_1,{\bf e}_2)\equiv \frac{1}{2}\biggl[Y_{\ell _1}^{m_1}({\bf e}_1)
Y_{\ell _2}^{m_2*}({\bf e}_2)+Y_{\ell _1}^{m_1*}({\bf e}_1)Y_{\ell _2}^{m_2}({\bf e}_2)\biggr] .
\end{equation}
$R^{\alpha }_{\rm R}({\bf e}_1,{\bf e}_2)$ 
is not symmetric under a permutation of the two directions. It
satisfies $R^{\alpha}_{\rm R}({\bf e}_2,{\bf e}_1)=
R^{\bar{\alpha }}_{\rm R}({\bf e}_1,{\bf e}_2)$, 
where the index $\bar{\alpha } $ is defined by 
$\bar\alpha\equiv
{\tiny\{\begin{array}{ccc}
         \! \ell_2 \! & \! \ell_1 \! \\
         \! m_2    \! & \! m_1    \! 
                           \end{array}\}}$. 
The symmetries in the indices and in the directions will play 
an important r\^ole in what follows. Using the previous 
definitions, we can introduce a quantity which is 
symmetric {\it both} under a permutation of the two directions 
{\it and} under a permutation of the columns of internal 
indices in $\alpha$ 
\begin{equation}
\label{defRS2}
R^{\alpha }_{\rm S}({\bf e}_1,{\bf e}_2)\equiv \frac{1}{4}\biggl[Y_{\ell _1}^{m_1}({\bf e}_1)
Y_{\ell _2}^{m_2*}({\bf e}_2)+Y_{\ell _1}^{m_1}({\bf e}_2)Y_{\ell _2}^{m_2*}({\bf e}_1)+
Y_{\ell _1}^{m_1*}({\bf e}_1)
Y_{\ell _2}^{m_2}({\bf e}_2)+Y_{\ell _1}^{m_1*}({\bf e}_2)Y_{\ell _2}^{m_2}({\bf e}_1)\biggr],
\end{equation}
i.e. we have
$R^{\alpha }_{\rm S}({\bf e}_1,{\bf e}_2) = 
R^{\alpha }_{\rm S}({\bf e}_2,{\bf e}_1) = 
R^{\bar\alpha }_{\rm S}({\bf e}_1,{\bf e}_2) = 
R^{\bar\alpha }_{\rm S}({\bf e}_2,{\bf e}_1)$. 
Finally, we will also employ a symmetrized 
Kr\"onecker symbol defined according to
\begin{equation}
\label{defdelta}
\delta ^{\alpha \alpha '}_{\rm S}\equiv \frac{1}{2}
\biggl(\delta _{\ell_1\ell_1'}\delta _{\ell_2\ell_2'}
\delta _{m_1m_1'}\delta _{m_2m_2'}+
\delta _{\ell_1\ell_2'}\delta _{\ell_2\ell_1'}
\delta _{m_1m_2'}\delta _{m_2m_1'}\biggr),
\end{equation} 
which is left unchanged under a permutation of the indices $\alpha $ and 
$\alpha '$ and also under permutations of the columns of each 
collective index separately.

Let us now turn to the computation of the best unbiased estimator; the
general definition of an estimator can be expressed as
\begin{equation}
\label{defEstH}
{\cal E}({\cal C}_{\alpha })\equiv \int \int {\rm d}\Omega _1 {\rm
d}\Omega _2 E^{\alpha }({\bf e}_1,{\bf e}_2)\Delta({\bf
e}_1)\Delta({\bf e}_2).
\end{equation}
The quantity ${\cal C}_{\alpha }$ is unchanged if we permute the 
columns of indices in $\alpha$, ${\cal C}_{\alpha } = {\cal
C}_{\bar\alpha }$ with $\bar\alpha$ given above. Being an estimator 
for ${\cal C}_{\alpha }$ it is then natural to assume that 
${\cal E}({\cal C}_{\alpha })$ possesses the same property, 
${\cal E}({\cal C}_{\alpha })={\cal E}({\cal C}_{\bar \alpha })$. 
Looking at the definition of $E^\alpha(\go,\gt)$, 
Eq. (\ref{defEstH}), we easily see that the 
weight function will also satisfy 
$E^{\alpha}(\go,\gt) = E^{\bar\alpha}(\go,\gt)$. Moreover, we take 
$E^{\alpha }({\bf e}_1,{\bf e}_2)$ symmetric under a permutation 
in the directions
${\bf e}_1$ and ${\bf e}_2$, i.e. 
$E_{\rm S}^{\alpha }({\bf e}_1,{\bf e}_2)=
E_{\rm S}^{\alpha }({\bf e}_2,{\bf e}_1)$. 

Now, we require that the estimator ${\cal E}(C_{\alpha })$ be
unbiased, which implies that the following relation must be fulfilled
\begin{equation}
\label{unbiH}
\int \int {\rm d}\Omega _1 {\rm d}\Omega _2 
E_{\rm S}^{\alpha }({\bf e}_1,{\bf e}_2)R^{\alpha '}
({\bf e}_1,{\bf e}_2)=\delta ^{\alpha \alpha '}_{\rm S}.
\end{equation}
We have required the presence of $\delta ^{\alpha \alpha '}_{\rm S}$
in the right hand side of the previous equation to respect the symmetries
in $\alpha$ of the weight function. In the previous equation the 
$E_{\rm S}^{\alpha }$ and $R^{\alpha '}$ are {\it a priori} complex. 
However, one can show that it is possible to work only with a real
weight function, 
$E_{\rm S}^{{\alpha }*}({\bf e}_1,{\bf e}_2)=E_{\rm S}^{\alpha }({\bf
e}_1,{\bf e}_2)$ and with the 
$R^{\alpha }_{\rm R}({\bf e}_1,{\bf e}_2)$ defined above:
the imaginary contributions would just increase the variance. 
(One could have also chosen to work with a pure imaginary weight function
and the imaginary part of $R^{\alpha }$). 
Therefore, the constraint can be written as 
\begin{equation}
\label{unbiHreal}
\int \int {\rm d}\Omega _1 {\rm d}\Omega _2 
E_{\rm S}^{\alpha }({\bf e}_1,{\bf e}_2)R_{\rm R}^{\alpha '}
({\bf e}_1,{\bf e}_2)=\delta ^{\alpha \alpha '}_{\rm S} ,
\end{equation}
where $E_{\rm S}^{\alpha }$ has been taken real.
Let us now calculate the variance of the estimator 
${\cal E}({\cal C}_{\alpha })$. Using Eqns. (\ref{4a}), 
we easily find that 
\begin{equation}
\label{varEH}
\sigma ^2_{{\cal E}({\cal C}_{\alpha })}=
2\int \int \int \int {\rm d}\Omega _1 {\rm d}\Omega _2{\rm d}\Omega _3 {\rm d}\Omega _4
E_{\rm S}^{\alpha }({\bf e}_1,{\bf e}_2)
E_{\rm S}^{\alpha }({\bf e}_3,{\bf e}_4)
\xi ({\bf e}_1\!\cdot\!{\bf e}_3)\xi ({\bf e}_2 \!\cdot\! {\bf e}_4).
\end{equation}
Our next step is to minimize this variance under the constraint given 
in Eq. (\ref{unbiHreal}). For this purpose, we introduce a set of Lagrange 
multipliers $\lambda ^{\alpha }$ and require that
\begin{equation}
\label{minivarH}
\delta \biggr[\sigma ^2_{{\cal E}({\cal C}_{\alpha })}+
\sum _{\alpha '}\lambda ^{\alpha }_{\alpha '}
\biggl(\int \int {\rm d}\Omega _1 {\rm d}\Omega _2 
E_{\rm S}^{\alpha }({\bf e}_1,{\bf e}_2)R^{\alpha '}_{\rm R}
({\bf e}_1,{\bf e}_2)-\delta ^{\alpha \alpha '}_{\rm S}\biggr)\biggr]=0 .
\end{equation}
At this point the precise meaning of the variation symbol 
$\delta$ matters. 
Before performing the variation, let us recall that the symmetries of
the weight function must be respected; hence we will have
\begin{equation}
\label{defvar}
\frac{{\rm \delta }E_{\rm S}^{\alpha }({\bf e}_i,{\bf e}_j)}
{{\rm \delta }E_{\rm S}^{\beta  }({\bf e}_k,{\bf e}_{\ell})}
=\frac{1}{2}\biggl[
{\delta({\bf e}_i\cd {\bf e}_k - 1)\over 2\pi}
{\delta({\bf e}_j\cd {\bf e}_{\ell } - 1)\over 2\pi}+
{\delta({\bf e}_i\cd {\bf e}_{\ell }- 1)\over 2\pi}
{\delta({\bf e}_j\cd {\bf e}_k - 1)\over 2\pi}
\biggr]\delta^{\alpha \beta}_{\rm S}.
\end{equation}
The $2\pi$'s in the denominators come from the fact that, 
while the direction ${\bf e}_i$ is expressed in terms of the corresponding 
spherical angles like ${\bf e}_i \equiv (\theta_{i}, \varphi_{i})$
and then ${\rm d}\Omega _i= {\rm d}\cos\theta_{i} {\rm d}\varphi_{i}$, 
there is an extra $2\pi$ factor in 
$\delta({\bf e}_i\cd {\bf e}_k - 1) = 2\pi \, 
\delta(\cos\theta_{i} - \cos\theta_{k}) \delta(\varphi_{i} - \varphi_{k})$. As a result of the variation, we obtain the following equation
\begin{equation}
\label{resuvar}
\int \int {\rm d}\Omega _1 {\rm d}\Omega _2
E_{\rm S}^{\alpha }({\bf e}_1,{\bf e}_2)\xi ({\bf e}_1\!\cdot\!{\bf e}_3)
\xi ({\bf e}_2\!\cdot\!{\bf e}_4)=\frac{1}{4}\sum _{\alpha '}\lambda ^{\alpha }_{\alpha '}
R_{\rm S}^{\alpha '}({\bf e}_3,{\bf e}_4).
\end{equation}
The quantity $R_{\rm S}^{\alpha '}({\bf e}_3,{\bf e}_4)$ appears
naturally as a result of Eq. (\ref{defvar}). The previous equation
should be compared with Eq. (\ref{resumini}) of the previous
subsection. The result of the variation is a relation between the
weight function and the Lagrange multiplier. Our aim now is to get an
explicit expression for the weight function.  This can be done by
using the inverse two-point correlation function $\xi^{-1}$ which
satisfies (see also the Appendix)
\be 
\label{invve}
\int {\rm d}\Omega _j \xi ({\bf
e}_i\!\cdot\!{\bf e}_j)\xi ^{-1}({\bf e}_j\!\cdot\!{\bf e}_k)\equiv \delta
({\bf e}_i \cd {\bf e}_k-1).  
\ee 
In the case of the three-point correlator, this definition leads to
subtleties which will be examined in detail in the next section. 
Multiplying Eq. (\ref{resuvar}) by 
$\xi ^{-1}({\bf e}_{1'}\cdot{\bf e}_3)
 \xi ^{-1}({\bf e}_{2'}\cdot{\bf e}_4)$, integrating over
directions ${\bf e}_3$ and ${\bf e}_4$ and relabelling indices, 
we arrive at
\begin{equation}
\label{weight2H}
E_{\rm S}^{\alpha }({\bf e}_1,{\bf e}_2)=\frac{1}{4}
\frac{1}{(2\pi)^2}
\sum _{\alpha '}\int \int {\rm d}\Omega _3 {\rm d}\Omega _4
\lambda ^{\alpha }_{\alpha '}
\xi ^{-1}({\bf e}_1\!\cdot\!{\bf e}_3)
\xi ^{-1}({\bf e}_2\!\cdot\!{\bf e}_4)
R_{\rm S}^{\alpha '}({\bf e}_3,{\bf e}_4).
\end{equation}
This equation for the second moment is the analogous of Eq. (21) of
(\cite{Hea98}) obtained for the third moment.  It is clear from the
previous section that, at this stage, our final goal has not yet been
reached. The weight function is still expressed in terms of the
Lagrange multipliers. The correct way to proceed is to remove the
latter using the constraint given by Eq. (\ref{unbiHreal}) as it was
done in the previous subsection. Then, we first multiply
Eq. (\ref{weight2H}) by the quantity $R_{\rm S}^{\alpha ''}({\bf
e}_1,{\bf e}_2)$ and, after having integrated over solid angles
$\Omega _1$ and $\Omega_2$, this leads to the equation that the
Lagrange multipliers must satisfy
\begin{equation}
\label{LH2}
16{\cal C}_{\ell_1''}{\cal C}_{\ell_2''}
\delta ^{\alpha \alpha ''}_{\rm S}
=
\biggl[
(-1)^{m_1''+m_2''}
\lamudeux{\ell_1''}{\ell_2''}{-m_1''}{-m_2''}{\alpha}
+
(-1)^{m_1''+m_2''}
\lamudeux{\ell_2''}{\ell_1''}{-m_2''}{-m_1''}{\alpha}
+
\lamudeux{\ell_1''}{\ell_2''}{m_1''}{m_2''}{\alpha}
+
\lamudeux{\ell_2''}{\ell_1''}{m_2''}{m_1''}{\alpha}
\biggr] .
\end{equation}
This equation is the analogue of the equation $\lambda ^{\ell }_{\ell
_1}=-4[{\cal C}_{\ell _1}^2/(2\ell _1+1)] \delta ^{\ell }_{\ell _1}$
of the previous subsection. Here, the difference is that a {\em
combination} of Lagrange multipliers with different indices appears
rather than the Lagrange multiplier itself. 
However, we can reconstruct exactly this combination in the right 
hand side of Eq. (\ref{weight2H}). 
Indeed, we just have to multiply each side
of Eq. (\ref{LH2}) by $R_{\rm S}^{\alpha ''}({\bf e}_1,{\bf
e}_2)$ and perform the sum over $\alpha ''$. Using the symmetry 
properties of $R_{\rm S}^{\alpha ''}({\bf e}_1,{\bf e}_2)$ we obtain 
\begin{equation}
\label{lRH}
4{\cal C}_{\ell_1}{\cal C}_{\ell_2}
R_{\rm S}^{\alpha }({\bf e}_1,{\bf e}_2)=
\sum _{\alpha ''}\lambda ^{\alpha }_{\alpha ''}
R_{\rm S}^{\alpha ''}({\bf e}_1,{\bf e}_2).
\end{equation}
We now insert this equation in the right hand side of
Eq. (\ref{weight2H}) to remove the Lagrange multipliers and find
\begin{equation}
\label{bestEH}
E^{\alpha }_{\rm S , Best}({\bf e}_1,{\bf e}_2)=\frac{1}{4\pi ^2}
{\cal C}_{\ell_1}{\cal C}_{\ell_2}
\int \int {\rm d}\Omega _3 {\rm d}\Omega _4
\xi ^{-1}({\bf e}_1\!\cdot\!{\bf e}_3)
\xi ^{-1}({\bf e}_2\!\cdot\!{\bf e}_4)
R_{\rm S}^{\alpha }({\bf e}_3,{\bf e}_4).
\end{equation}

To go further, we express $\xi^{-1}$ in Eq. (\ref{bestEH}) explicitly
in terms of spherical harmonics (see the Appendix) which yields
$E_{\rm S , Best}^{\alpha }({\bf e}_1,{\bf e}_2) =R_{\rm S}^{\alpha
}({\bf e}_1,{\bf e}_2)$.
Now, we just plug this result into Eq. (\ref{defEstH}) to 
get the final explicit expression of the best unbiased
estimator ${\cal E}_{\rm Best}({\cal C}_{\alpha })$
\begin{equation}
\label{exEH}
{\cal E}_{\rm Best}({\cal C}_{\alpha })=\frac{1}{2}\biggl(a_{\ell_1}^{m_1*}a_{\ell_2}^{m_2}
+a_{\ell_1}^{m_1}a_{\ell_2}^{m_2*}\biggr) .
\end{equation}
Some remarks are in order here. First, despite appearances, one {\em
cannot} deduce ${\cal E}_{\rm Best}({\cal C}_{\alpha })$ from ${\cal
E}_{\rm Best}({\cal C}_{\ell_1})$. It is clear that the following
equation holds
\begin{equation}
\label{rel}
\langle {\cal E}({\cal C}_{\alpha })\rangle 
=\langle {\cal E}({\cal C}_{\ell_1})\rangle  \delta _{\ell_1\ell_2}\delta _{m_1m_2}.
\end{equation}
However, from this equation we are {\em not} allowed to conclude that
${\cal E}_{\rm Best}({\cal C}_{\alpha }) ={\cal E}_{\rm Best}({\cal
C}_{\ell_1})\delta _{\ell_1\ell_2}\delta _{m_1m_2}$. Therefore, knowing
one of the best unbiased estimators does not allow us to infer the
other best one.
To be specific, if one assumed the previous wrong relation, 
from Eq. (\ref{estmulti}) one would get 
\begin{equation}
\label{EHbestwrong}
{\cal E}_{\rm Best}({\cal C}_{\alpha })=\frac{1}{2\ell_1+1}\biggl(\sum
_{m=-\ell_1}^{\ell_1} a_{\ell_1}^{m*}a_{\ell_1}^{m}\biggr)\delta
_{\ell_1\ell_2}\delta _{m_1m_2} \quad \mbox{(false)} ,
\end{equation}
and although this estimator is unbiased, it is {\em not} the best
one. 
The second remark is that although the estimator given by Eq. (\ref{exEH}) 
could be naively regarded as a trivial one, this is not the case.
Indeed, the estimator of
Eq. (\ref{EHbestwrong}) is as trivial as the actual best one.
The moral is then that there exist simple choices which lead to
the wrong answer. The only reliable method in the problem of
minimizing the variance is therefore the one exposed above.

\section{Three-point correlators}

\subsection{Best estimator for the angular bispectrum ${\cal C}_{\ell _1 \ell_2 \ell_3}$}


In this section, our aim is to determine the best unbiased estimator
for the angular bispectrum ${\cal C}_{\ell _1 \ell_2 \ell_3}$ defined
in the third of Eqns. (\ref{propa}). According to our general
prescription, the most general definition reads
\begin{equation}
\label{defE3}
{\cal E}({\cal C}_{\ell _1 \ell_2 \ell_3})\equiv \int \int \int {\rm
d}\Omega _1 {\rm d}\Omega _2 {\rm d}\Omega _3 E_{\rm S}^{\ell _1 \ell_2
\ell_3}({\bf e}_1,{\bf e}_2,{\bf e}_3) \Delta({\bf e}_1)\Delta ({\bf
e}_2)\Delta({\bf e}_3).
\end{equation}
As in the case of the two-point correlators, the weight function also
possesses the properties of being real and symmetric under arbitrary
permutations of directions ${\bf e}_i$.  
In addition, like ${\cal C}_{\ell _1 \ell_2 \ell_3}$, the weight function 
satisfies 
$E_{\rm S}^{\ell _1 \ell_2 \ell_3} = E_{\rm S}^{\ell _2 \ell_1 \ell_3}$, 
as well as for any other arbitrary permutation of the indices $\ell_i$.
We follow similar steps as
for the angular spectrum and therefore we choose to expand the weight
function on the basis of the spherical harmonics. Then, as in
Eq. (\ref{expomC_l}), we write
\begin{equation}
\label{defd3}
E_{\rm S}^{\ell _1 \ell_2 \ell_3}({\bf e}_1,{\bf e}_2,{\bf e}_3)=
\sum _{\ell_1' m_1'}
\sum _{\ell_2' m_2'}
\sum _{\ell_3' m_3'}
\coefdd{\ell_1'}{\ell_2'}{\ell_3'}{m_1'}{m_2'}{m_3'}
       {\ell_1}{\ell_2}{\ell_3}   
Y_{\ell_1'}^{m_1'}({\bf e}_1)
Y_{\ell_2'}^{m_2'}({\bf e}_2)Y_{\ell_3'}^{m_3'}({\bf e}_3).
\end{equation}
The properties of the weight function imply that the coefficients $d$ 
must satisfy equations similar to those given in Eqns. (\ref{propd})
\begin{equation}
\label{propd3}
\coefdd{\ell_1'}{\ell_2'}{\ell_3'}{m_1'}{m_2'}{m_3'}
       {\ell_1}{\ell_2}{\ell_3 *}
= (-1)^{m_1'+m_2'+m_3'}
\coefdd{\ell_1'}{\ell_2'}{\ell_3'}{-m_1'}{-m_2'}{-m_3'}
       {\ell_1}{\ell_2}{\ell_3}  
\quad
, 
\quad 
\coefdd{\ell_1'}{\ell_2'}{\ell_3'}{m_1'}{m_2'}{m_3'}
{\ell_1}{\ell_2}{\ell_3}
=
\coefdd{\ell_2'}{\ell_1'}{\ell_3'}{m_2'}{m_1'}{m_3'}
{\ell_1}{\ell_2}{\ell_3},
\end{equation}
where the last relation is in fact valid for arbitrary permutations of
any two columns of the collective subindex.  
Like the weight function, $d$ is also left invariant under 
arbitrary permutations of indices $\ell_i$ (not primed).
The estimator can be
expressed in terms of the coefficients $d$ and the $a_{\ell }^m$'s
only: inserting the expansion of the weight function in the
above expression for the estimator and using standard properties of the
spherical harmonics one obtains
\begin{equation}
\label{esti3ad}
{\cal E}({\cal C}_{\ell _1 \ell_2 \ell_3})
=
\sum _{\ell_1' m_1'}
\sum _{\ell_2' m_2'}
\sum _{\ell_3' m_3'}
\coefdd{\ell_1'}{\ell_2'}{\ell_3'}{m_1'}{m_2'}{m_3'}
       {\ell_1}{\ell_2}{\ell_3 *}
a_{\ell _1'}^{m_1'}a_{\ell _2'}^{m_2'}a_{\ell _3'}^{m_3'} .
\end{equation}
In practice, CMB observational settings are devised such that both the 
monopole and the dipole are subtracted from the anisotropy maps. 
This means that the coefficients $d$ in the last equation are only 
non-vanishing for indices $\ell_i' \ge 2$ in the collective subindex.  
Moreover, the coefficients $d$ satisfy 
$\ell _1 +\ell _2 +\ell _3=\mbox{even}$.
We must now require that our general estimator given by
Eq. (\ref{esti3ad}) be unbiased, i.e. $\langle {\cal E}({\cal C}_{\ell
_1 \ell_2 \ell_3})\rangle ={\cal C}_{\ell _1 \ell_2 \ell_3}$. This
forces the coefficients $d$ to fulfill the following constraint 
\begin{equation}
\label{cons3}
\sum _{m_1'm_2'm_3'}
\coefdd{\ell_1'}{\ell_2'}{\ell_3'}{m_1'}{m_2'}{m_3'}{\ell _1}{\ell _2}{\ell _3 *}
\wjma{\ell _1'}{\ell _2'}{\ell _3'}{m_1'}{m_2'}{m_3'}
= \delta _{\rm S}^{\ell_{i} \ell_{j}'},
\end{equation}
where we have defined a new symmetrized Kr\"onecker symbol, this
time for the $\ell$ multipole indices only, as follows 
\be 
\delta _{\rm S}^{\ell_{i} \ell_{j}'} \equiv \frac{1}{6} \biggl(
\delta_{\ell_1\ell_1'}\delta_{\ell_2\ell_2'}\delta_{\ell_3\ell_3'}
+\mbox{ 5 additional permutations }  \biggr). 
\ee 
It is easy to check that the constraint equation satisfies the
conditions imposed by Eqns. (\ref{propd3}) on the coefficients $d$. In
particular, let us justify the presence of the symbol $\delta _{\rm
S}^{\ell_{i} \ell_{j}'}$.  Using the previous properties for $d$,
relabelling the indices $m_1' \leftrightarrow m_2'$ in
Eq. (\ref{cons3}) and finally noting that $\ell_1'+\ell_2'+\ell_3'=
\ell_1+\ell_2+\ell_3=\mbox{even}$, which allows us to permute any two
columns of the Wigner 3$j$-symbol, one verifies that the left hand
side of the constraint is invariant under
$\ell_1'\leftrightarrow\ell_2'$.  The same applies for any pair of
$\ell$ multipole indices and this explains the presence of the
symmetrized $\delta_{\rm S}^{\ell_{i} \ell_{j}'}$ in
Eq. (\ref{cons3}).  We see from this that all coefficients $d$ that do
not satisfy $\ell_1'+\ell_2'+\ell_3'=\mbox{even}$ do not enter the
constraint.  We will show below that these terms only increase the
variance and as a consequence one can take them equal to zero. In
particular, note that this is the case for a coefficient $d$ with
$\ell_1'=\ell_2'$ and $\ell_3' = \mbox{odd}$. This property will turn
out to be useful in what follows.

We are now in a position to calculate the variance of the
estimator. Looking at Eq. (\ref{esti3ad}) we see that this requires
the computation of the sixth moment of the $\almn$'s, see
Eq. (\ref{6a}). After having made use of the properties of the
coefficients $d$ and rearranging the resulting 15 terms into two groups,
straightforward algebra yields
\begin{equation}
\label{var3} 
\langle \left[ {\cal E}({\cal C}_{\ell _1 \ell_2 \ell_3})\right]^2 \rangle
=
\sum _{\ell_1' m_1'}
\sum _{\ell_2' m_2'}
\sum _{\ell_3' m_3'}
{\cal C}_{\ell_1'}{\cal
C}_{\ell_2'}{\cal C}_{\ell_3'} \biggl[6
\coefdd{\ell_1'}{\ell_2'}{\ell_3'}{m_1'}{m_2'}{m_3'}{\ell _1}{\ell
_2}{\ell _3 *}
\coefdd{\ell_1'}{\ell_2'}{\ell_3'}{m_1'}{m_2'}{m_3'}{\ell _1}{\ell
_2}{\ell _3} 
+9(-1)^{m_1'+m_2'}
\coefdd{\ell_1'}{\ell_1'}{\ell_3'}{m_1'}{-m_1'}{m_3'}{\ell _1}{\ell
_2}{\ell _3 *}
\coefdd{\ell_2'}{\ell_2'}{\ell_3'}{m_2'}{-m_2'}{m_3'}{\ell _1}{\ell
_2}{\ell _3} \biggr] .
\end{equation}
The square of the variance of ${\cal E}({\cal C}_{\ell _1 \ell_2 \ell_3})$ 
is given by
\be
\label{vari2}
\sigma ^2 _{{\cal E}({\cal C}_{\ell _1 \ell_2 \ell_3})}
=
\la \left[ {\cal E}({\cal C}_{\ell _1 \ell_2 \ell_3})\right]^2 \ra -
\la        {\cal E}({\cal C}_{\ell _1 \ell_2 \ell_3})  \ra^2 .
\ee
Following the discussion in \S 2, the term $\la \left[
{\cal E}({\cal C}_{\ell _1 \ell_2 \ell_3})\right]^2 \ra$ is of order
$\epsilon^0$ whereas the lowest non-vanishing order of $\la {\cal
E}({\cal C}_{\ell _1 \ell_2 \ell_3}) \ra^2$ is $\epsilon^2$.
Therefore, the latter one will not enter the minimization procedure and 
the variance squared will be written as 
$\sigma ^2 _{{\cal E}({\cal C}_{\ell _1 \ell_2
\ell_3})} \approx \la \left[ {\cal E}({\cal C}_{\ell _1 \ell_2
\ell_3})\right]^2 \ra$. However, let us notice that this does not occur
in the case of the two-point correlator. Indeed, as we have seen, in 
that case both terms contributing to the square of the variance are 
of the same order in $\epsilon$. Then, it follows that 
Eq. (\ref{var3}) corresponds to Eq. (\ref{varCl}) in the last section. 

Let us now examine the structure of the variance in more detail. The
term $6 d^* d \propto \Re^2(d)+\Im^2(d)$ in Eq. (\ref{var3}) is
analogous to the one in Eq. (\ref{varCl}).  However, here there is
another contribution, the $9 d^* d$ term, which will play a crucial
r\^ole in what follows.  The imaginary part of this term of course 
vanishes, as the variance must be real. However, the real part of 
it contains a contribution like 
\begin{equation}
\label{var3ca}
9 \sum_{\ell_3'}
{\cal C}_{\ell_3'}
\sum _{m_3'} 
\left[
\sum_{\ell_1'}{\cal C}_{\ell_1'}
\sum _{m_1'} 
(-1)^{m_1'}
\Im\left(
\coefdd{\ell_1'}{\ell_1'}{\ell_3'}
       {m_1'}{-m_1'}{m_3'}{\ell_1}{\ell_2}{\ell_3}
\right)
\right]
\left[
\sum_{\ell_2'}{\cal C}_{\ell_2'}
\sum _{m_2'} 
(-1)^{m_2'}
\Im\left(
\coefdd{\ell_2'}{\ell_2'}{\ell_3'}
       {m_2'}{-m_2'}{m_3'}{\ell_1}{\ell_2}{\ell_3}
\right)
\right]
=
9
\sum_{\ell_3'}
{\cal C}_{\ell_3'}
\sum _{m_3'}
\left[\ldots \right]^2,
\end{equation}
where the quantity $\left[\ldots \right]^2$ depends on the indices
$\ell_1 , \ell_2 , \ell_3$ and $\ell_3', m_3'$ and is strictly
positive or zero. Of course, there is a similar term coming from the
real part of $d$.  Thus, we see that the various contributions of the 
imaginary part
of the coefficients $d$ to the two terms, $6 d^* d$ and $9 d^* d$,
only increase the variance.  Since we know that a vanishing
imaginary part does satisfy the constraint Eq. (\ref{cons3}), 
it can be disregarded in the sequel.
Therefore,
Eq. (\ref{var3}) can then be written solely in terms of {\em real} 
coefficients $d$ as follows
\begin{equation}
\label{var3bis}
\sigma ^2 _{{\cal E}({\cal C}_{\ell _1 \ell_2 \ell_3})} 
=
\sum _{\ell_1' m_1'}
\sum _{\ell_2' m_2'}
\sum _{\ell_3' m_3'}
{\cal C}_{\ell_1'}{\cal
C}_{\ell_2'}{\cal C}_{\ell_3'} \biggl[6
\biggl(
\coefdd{\ell_1'}{\ell_2'}{\ell_3'}{m_1'}{m_2'}{m_3'}{\ell _1}{\ell
_2}{\ell _3 }
\biggr)^2
+9(-1)^{m_1'+m_2'}
\coefdd{\ell_1'}{\ell_1'}{\ell_3'}{m_1'}{-m_1'}{m_3'}{\ell _1}{\ell
_2}{\ell _3}
\coefdd{\ell_2'}{\ell_2'}{\ell_3'}{m_2'}{-m_2'}{m_3'}{\ell _1}{\ell
_2}{\ell _3} \biggr].
\end{equation}
Our next move now is to minimize this variance with respect to the
coefficients $d$, taking into account the constraint of
Eq. (\ref{cons3})
\begin{equation}
\label{mini3B}
\delta \biggl\{\sigma ^2 _{{\cal E}({\cal C}_{\ell _1 \ell_2 \ell_3})}
+\sum _{\ell _1'\ell _2'\ell _3'}
\lambda _{\ell _1'\ell _2'\ell _3'}^{\ell _1\ell _2\ell _3}
\biggl[
\sum _{m_1'm_2'm_3'}
\coefdd{\ell_1'}{\ell_2'}{\ell_3'}{m_1'}{m_2'}{m_3'}
       {\ell _1}{\ell_2}{\ell _3} 
\wjma{\ell _1'}{\ell _2'}{\ell _3'}{m_1'}{m_2'}{m_3'} -
\delta_{\rm S}^{\ell_{i} \ell_{j}'}
\biggr]\biggr\}=0 .
\end{equation}
This equation is the analogous to Eq. (\ref{mini}).  As before, one
needs to give a concrete meaning to the symbol ${\rm \delta }$ of the
variation. Its definition must respect the symmetries of the
coefficients $d$ and hence we take
\begin{equation}
\label{vard}
\frac{{\rm \delta }
\coefdd{\ell_1'}{\ell_2'}{\ell_3'}{m_1'}{m_2'}{m_3'}{\ell _1}{\ell _2}{\ell _3}}{
{\rm \delta }
\coefdd{\ell_1''}{\ell_2''}{\ell_3''}{m_1''}{m_2''}{m_3''}{\ell _1}{\ell _2}{\ell _3}}
=
\frac{1}{6}\biggl\{\frac{1}{2}\delta _{\ell_1'\ell_1''}\delta _{\ell_2'\ell_2''}
\delta _{\ell_3'\ell_3''}\biggl[
\delta _{m_1'm_1''}\delta _{m_2'm_2''}\delta _{m_3'm_3''}
+(-1)^{m_1'+m_2'+m_3'}
\delta _{m_1'-m_1''}\delta _{m_2'-m_2''}\delta _{m_3'-m_3''}\biggr] +
\mbox{5 terms}\biggr\} .
\end{equation}
Then, Eq. (\ref{mini3B}) leads to 
\begin{eqnarray}
\label{mini3resu}
&&12{\cal C}_{\ell _1'}{\cal C}_{\ell _2'}{\cal C}_{\ell _3'}
\coefdd{\ell_1'}{\ell_2'}{\ell_3'}{m_1'}{m_2'}{m_3'}{\ell _1}{\ell
_2}{\ell _3} + \lambda _{\ell _1'\ell _2'\ell _3'}^{\ell _1\ell _2\ell
_3} \wjma{\ell _1'}{\ell _2'}{\ell _3'}{m_1'}{m_2'}{m_3'}
+ 6(-1)^{m_2'}{\cal C}_{\ell_2'}{\cal C}_{\ell_3'}
\delta_{\ell_1'\ell_2'}\delta_{m_1'-m_2'} 
\sum _{\ell m}{\cal C}_{\ell} (-1)^{m} 
\coefdd{\ell}{\ell}{\ell_3'}{\ m \ }{\ -m \ }{\ m_3' \ }
{\ell_1 }{\ell _2 }{\ell _3 } 
\nonumber \\ &+& 
6(-1)^{m_3'}{\cal C}_{\ell_3'}{\cal C}_{\ell_1'}
\delta_{\ell_2'\ell_3'}\delta_{m_2'-m_3'} 
\sum _{\ell m}{\cal C}_{\ell} (-1)^{m} 
\coefdd{\ell}{\ell}{\ell_1'}{\ m \ }{ \ -m \ }{ \ m_1' \ }
{\ell_1}{\ell _2}{\ell _3} 
+ 6(-1)^{m_1'}{\cal C}_{\ell_1'}{\cal C}_{\ell_2'}
\delta_{\ell_3'\ell_1'}\delta_{m_3'-m_1'} 
\sum _{\ell m}{\cal C}_{\ell} (-1)^{m} 
\coefdd{\ell}{\ell}{\ell_2'}{ \ m \ }{ \ -m \ }{ \ m_2' \ }
{\ell_1}{\ell _2}{\ell _3} 
=0 .
\end{eqnarray}
This formula, together with Eq. (\ref{cons3}), form a set of
equations which completely determines the best unbiased estimator. We
see the complicated structure of these equations. The last
three terms come from the $9 d d$ term in the
variance and are not present in the case of the angular
spectrum. 

{}From this last equation and using the constraint Eq. (\ref{cons3}) 
we can get the general expression for the Lagrange multipliers. 
Thus, we multiply Eq. (\ref{mini3resu}) by the appropriate 3$j$-symbol
and we sum over the three indices $m_i'$. The first term is exactly the 
constraint and produces a $\delta_{\rm S}^{\ell_{i} \ell_{j}'}$.
Using the fact that a triple sum over the $m_i$'s of the squared
of a 3$j$-symbol gives unity, the second term yields the Lagrange 
multipliers themselves.
Finally, the last three terms vanish: indeed, after straightforward 
manipulations one generates a term like 
\be
\label{oned}
\coefdd{\ell}{\ell}{\ell_3'}{m}{-m}{0}{\ell_1}{\ell_2}{\ell_3}
\sum_{m_1'} (-1)^{m_1'}
\wjma{\ell _1'}{\ell _1'}{\ell _3'}{m_1'}{-m_1'}{0}
\ee
for the third term of Eq. (\ref{mini3resu}) and analogously for the last
two ones. As we mentioned previously, the coefficient $d$ in 
Eq. (\ref{oned}) vanishes unless 
$\ell_3'$ is even. In this case, recalling the identity (\cite{molle})
\be
\label{magicJ}
\sum_{m_1'} (-1)^{m_1'}
\wjma{\ell _1'}{\ell _1'}{\ell _3'}{m_1'}{-m_1'}{0}
= (-1)^{\ell _1'} \sqrt{2{\ell _1'}+1} \; \delta_{{\ell _3'} 0} ,
\quad {\ell _3'} = \mbox{even} ,
\ee
we see that there will only be a non-vanishing term if $\ell_3' = 0$.
But, the corresponding $d$ is zero because $\ell_3' < 2$ and therefore 
terms of this kind do not contribute to the Lagrange multipliers. 
Then, these are given by
\begin{equation}
\label{LagMul3}
\lambda _{\ell _1'\ell _2'\ell _3'}^{\ell _1\ell _2\ell _3}=
-12{\cal C}_{\ell _1'}{\cal C}_{\ell _2'}{\cal C}_{\ell _3'}
\delta_{\rm S}^{\ell_{i} \ell_{j}'} .
\end{equation}
Plugging this into Eq. (\ref{mini3resu}), one has
\begin{eqnarray}
\label{mini3resubis}
&&12{\cal C}_{\ell _1'}{\cal C}_{\ell _2'}{\cal C}_{\ell _3'}
\biggl[
\coefdd{\ell_1'}{\ell_2'}{\ell_3'}{m_1'}{m_2'}{m_3'}{\ell 
_1}{\ell_2}{\ell _3} 
-
\delta_{\rm S}^{\ell_{i} \ell_{j}'}
\wjma{\ell _1'}{\ell _2'}{\ell _3'}{m_1'}{m_2'}{m_3'}
\biggr]
+ 6(-1)^{m_2'}{\cal C}_{\ell_2'}{\cal C}_{\ell_3'}
\delta_{\ell_1'\ell_2'}\delta_{m_1'-m_2'} 
\sum _{\ell m}{\cal C}_{\ell} (-1)^{m} 
\coefdd{\ell}{\ell}{\ell_3'}{ \ m \ }{ \ -m \ }{ \ m_3' \ }
{\ell_1}{\ell _2}{\ell _3} 
\nonumber \\ &+& 
6(-1)^{m_3'}{\cal C}_{\ell_3'}{\cal C}_{\ell_1'}
\delta_{\ell_2'\ell_3'}\delta_{m_2'-m_3'} 
\sum _{\ell m}{\cal C}_{\ell} (-1)^{m} 
\coefdd{\ell}{\ell}{\ell_1'}{ \ m \ }{ \ -m \ }{ \ m_1' \ }
{\ell_1}{\ell _2}{\ell _3} 
+ 6(-1)^{m_1'}{\cal C}_{\ell_1'}{\cal C}_{\ell_2'}
\delta_{\ell_3'\ell_1'}\delta_{m_3'-m_1'} 
\sum _{\ell m}{\cal C}_{\ell} (-1)^{m} 
\coefdd{\ell}{\ell}{\ell_2'}{ \ m \ }{ \ -m \ }{ \ m_2' \ }
{\ell_1}{\ell _2}{\ell _3} 
=0 .
\eea
This is the final equation to be solved in order to determine the 
best unbiased estimator. A solution is 
\begin{equation}
\label{sold3}
\coefdd{\ell_1'}{\ell_2'}{\ell_3'}{m_1'}{m_2'}{m_3'}{\ell_1 }{\ell_2 }{\ell_3 }
=\wjma{\ell _1'}{\ell _2'}{\ell _3'}{m_1'}{m_2'}{m_3'}
\delta_{\rm S}^{\ell_{i} \ell_{j}'} .
\end{equation}
This leads to
\begin{equation}
\label{solest3}
{\cal E}_{\rm Best}({\cal C}_{\ell_1 \ell_2 \ell_3 })
= \sum _{m_1' m_2' m_3'}
\wjma{\ell_1 }{\ell_2 }{\ell_3 }{m_1'}{m_2'}{m_3'}
a_{\ell_1 }^{m_1'}a_{\ell_2 }^{m_2'}a_{\ell_3 }^{m_3'} .
\end{equation}
This is the main result of this subsection. 

Given that we now know the best unbiased estimator for ${\cal
C}_{\ell_1 \ell_2 \ell_3 }$, one can compute its variance, the
smallest one amongst all possible estimator variances. In 
(\cite{GanMar00}) we have already calculated it and reads
\be
\label{vari2gm}
\sigma ^2 _{{\cal E}_{\rm Best}({\cal C}_{\ell _1 \ell_2 \ell_3})}
=
{\cal C}_{\ell_1} {\cal C}_{\ell_2} {\cal C}_{\ell_3}
(1+\delta_{\ell_1\ell_2}+\delta_{\ell_2\ell_3}+\delta_{\ell_3\ell_1}
+ 2 ~ \delta_{\ell_1\ell_2}\delta_{\ell_2\ell_3}) .
\ee
In the same reference a plot of this variance 
for low order multipoles can 
also be found. This is what one could dub (the square of) the 
`bispectrum cosmic variance' in perfect analogy with
$\sigma^2_{{\cal E}_{\rm Best}({\cal C}_{\ell})}
= 2 {\cal C}_{\ell}^2 / (2\ell+1)$, which is (the square of) the 
variance of the best unbiased estimator for the angular spectrum,
commonly known as the `cosmic variance'. 

Let us conclude this subsection by comparing our results with those
recently appeared in the literature. An estimator restricted to the
diagonal case $\ell _1=\ell _2=\ell _3$ has been proposed in
(\cite{Feretal98}, see also \cite{newturn} for an extension of their
analysis) for ${\cal B}_{\ell }\equiv {\cal C}_{\ell \ell \ell}$ and
reads
\begin{equation}
\label{BFMG}
{\cal E}({\cal B}_{\ell })=\frac{1}{2\ell +1}
\wjma{\ell }{\ell }{\ell }{0}{0}{0}^{-3/2}
\sum _{m_1m_2m_3}
\wjma{\ell }{\ell }{\ell }{m_1}{m_2}{m_3}
a_{\ell }^{m_1}a_{\ell }^{m_2}a_{\ell }^{m_3}.
\end{equation}
In that work, the aim of the authors was not to seek the best
estimator, but to use Eq. (\ref{BFMG}) to analyse the non-Gaussian
features of the 4-yr COBE-DMR data.  It is easy to see that their
estimator does {\em not} satisfy the constraint (\ref{cons3}), i.e.
the estimator is biased. This is due to the presence of the overall
prefactor in front of the triple sum in Eq. (\ref{BFMG}).  However, as
we have proven above, getting rid of it produces the best unbiased
estimator ${\cal E}_{\rm Best}({\cal C}_{\ell_1 \ell_2 \ell_3 })$,
Eq. (\ref{solest3}).

\subsection{Best estimator for the third moment ${\cal B}^{\alpha }$}

We now seek an estimator for ${\cal B}^\alpha \equiv \bigl\langle
a_{\ell_1 }^{m_1} a_{\ell_2}^{ m_2} a_{\ell_3}^{ m_3} \bigr\rangle$
where, as in the last section, it is convenient to define
a collective index 
$\alpha\equiv 
{\tiny\{\begin{array}{ccc}
         \! \ell_1 \! & \! \ell_2 \!  & \! \ell_3 \\
         \! m_1    \! & \! m_2    \!  & \! m_3    
                           \end{array}\}}$.
This question has already been addressed in (\cite{Hea98}).
As we did with the second moment, our starting expression for an 
unbiased cubic estimator 
${\cal E}({\cal B}^{\alpha })$ of ${\cal B}^\alpha$ will be in the form
\be
{\cal E}({\cal B}^{\alpha }) = \int\int\int\dO{1}\dO{2}\dO{3}
E^\alpha(\go,\gt,\gth) \Delta(\go) \Delta(\gt) \Delta(\gth)
\label{gralesti}
\ee
The goal is to find the weight function $E^\alpha(\go,\gt,\gth)$ 
that minimizes the variance of the estimator.
As above, the quantity ${\cal B}^\alpha$ is unchanged if we permute arbitrary columns
of indices in $\alpha$:
${\cal B}^\alpha = {\cal B}^{\bar\alpha}$, where for instance
$\bar\alpha\equiv
{\tiny\{\begin{array}{ccc}
         \! \ell_2 \! & \! \ell_1 \!  & \! \ell_3 \\
         \! m_2    \! & \! m_1    \!  & \! m_3    
                           \end{array}\}}$.
${\cal E}({\cal B}^{\alpha })$ has the same properties as ${\cal
B}^\alpha$ and then it follows that ${\cal E}({\cal B}^{\alpha }) =
{\cal E}({\cal B}^{\bar\alpha })$ for any column-permutated
$\bar\alpha$.  This implies that the $E^\alpha(\go,\gt,\gth)$
satisfies $E^{\alpha}(\go,\gt,\gth) = E^{\bar\alpha}(\go,\gt,\gth)$.

Now, is $E^\alpha(\go,\gt,\gth)$ also symmetric under permutations in
the directions $\go,\gt,\gth$~? From its definition we cannot
know, for these directions are integrated over in the above defining
equation. Unlike the case for the second moment discussed before, here
$E^\alpha$ cannot be decomposed into a symmetric and antisymmetric parts.
However, we can always write $E^\alpha = E_{\rm S}^\alpha + \mbox{something}$,
and show that this last contribution to Eq. (\ref{gralesti}) vanishes.
Therefore, there is no loss of generality in working with 
$E_{\rm S}^\alpha(\go,\gt,\gth) \equiv {1\over 6}
[E^\alpha(\go,\gt,\gth)+ \mbox{ 5 terms }]$ 
which is symmetric under arbitrary permutations of directions ${\bf e}_i$.

Demanding the estimator ${\cal E}({\cal B}^{\alpha })$ to be unbiased,
$\la {\cal E}({\cal B}^{\alpha }) \ra = {\cal B}^\alpha$, yields the first 
constraint equation that the weight function $E_{\rm S}^\alpha(\go,\gt,\gth)$ 
must satisfy
\be
\int\int\int\dO{1}\dO{2}\dO{3}
E_{\rm S}^\alpha(\go,\gt,\gth) R_{\rm R}^{\alpha'}(\go,\gt,\gth)
= \delta_{\rm S}^{\alpha \alpha'} ,
\label{constraint}
\ee
where $R^{\alpha'}(\go,\gt,\gth)\equiv 
Y_{\ell_1'}^{m_1'}({\bf e}_1)
Y_{\ell_2'}^{m_2'}({\bf e}_2)
Y_{\ell_3'}^{m_3'}({\bf e}_3)$ 
and $R_{\rm R}^{\alpha'}(\go,\gt,\gth)$ 
is its real part. The form of $R^{\alpha}$
comes from the expression of $\xi_3$, viz. 
$\la \Delta(\go) \Delta(\gt) \Delta(\gth) \ra = 
\sum_{\alpha} {\cal B}^{\alpha} R^{\alpha}(\go,\gt,\gth)$.
The symmetrized Kr\"onecker symbol can be written as 
$\delta_{\rm S}^{\alpha \alpha'} \equiv 
{1\over 6}(\delta _{\ell_1 \ell_1'}
\delta _{m_1 m_1'}
\delta _{\ell_2 \ell_2'}
\delta _{m_2 m_2'}
\delta _{\ell_3 \ell_3'}
\delta _{m_3 m_3'}
+\mbox{ 5 terms})$ as required to
comply with the symmetry under permutations in the columns of
$\alpha$ in $E_{\rm S}^\alpha(\go,\gt,\gth)$.
In the above equation, $R_{\rm R}^{\alpha'}(\go,\gt,\gth)$
is clearly non-symmetric under a permutation of 
directions $\go,\gt,\gth$. 
However, as above, we can define a symmetrized combination
$R_{\rm S}^\alpha(\go,\gt,\gth) \equiv 
{1\over 6}[R_{\rm R}^\alpha(\go,\gt,\gth)+\mbox{5 terms}]$ 
(12 terms). Symmetrizing either in directions or in the 
columns of $\alpha$ in $R_{\rm R}^{\alpha}(\go,\gt,\gth)$ 
yields exactly the same $R_{\rm S}^\alpha(\go,\gt,\gth)$.
In the last equation and in what follows the weight function
$E_{\rm S}^\alpha$
is real for reasons similar to the ones exposed around 
Eq. (\ref{unbiHreal}) in the last section. 

Proceeding as in \S 3 and using Eq. (\ref{6a}), one gets
\be
\sigma^2_{{\cal E}({\cal B}^{\alpha })}
= \int\dO{1} \ldots \!\int\dO{6}
E_{\rm S}^\alpha(\go,\gt,\gth) 
E_{\rm S}^\alpha(\gf,\gfi,\gs) 
\xi(\go\cd\gf) \Big[6 \xi(\gt\cd\gfi)\xi(\gth\cd\gs)+
                      9 \xi(\gt\cd\gth)\xi(\gfi\cd\gs)\Big] ,
\label{bougnoula}
\ee
where, utilizing the symmetry of the coefficients $E_{\rm S}^\alpha(\gi,\gj,\gk)$ 
under ${\bf e}$-direction permutations,
only two types of $\xi$ products remain: first type, six terms where
all the three $\xi$'s mix directions of the first and
second $E_{\rm S}^\alpha$'s 
and, second type, nine terms where only one $\xi$ [in the above
equation, $\xi(\go\cd\gf)$] does it.

To minimize the remaining variance under the constraint 
(\ref{constraint}) we introduce a set of Lagrange multipliers
$\lambda^{\alpha}$ and write
\be
\delta
\Big[
\sigma^2_{{\cal E}({\cal B}^{\alpha })} - 
\sum_{\alpha'} \lambda^{\alpha}_{\alpha'} 
\Big(
\int\int\int\dO{1}\dO{2}\dO{3}
E_{\rm S}^\alpha(\go,\gt,\gth) 
R_{\rm R}^{\alpha'}(\go,\gt,\gth)
- \delta_{\rm S}^{\alpha \alpha'} 
\Big) 
\Big]=0 .
\label{mini2}
\ee

As already noted for the second moment, the symmetries of the weight
function must be respected in the variation; hence we have
\be
{\delta E_{\rm S}^\alpha(\go,\gt,\gth) \over \delta E_{\rm S}^\beta(\gi,\gj,\gk)}
= 
{1\over 6} \Big[
{\delta(\go\cd\gi - 1)\over 2\pi}
{\delta(\gt\cd\gj - 1)\over 2\pi}
{\delta(\gth\cd\gk - 1)\over 2\pi}
+ {\rm 5 ~terms}
\Big]  
\delta_{\rm S}^{\alpha \beta} .
\ee

Now, we vary Eq. (\ref{mini2}) and, after some algebra, we come up with
\be
2 \int\!\int\!\int\dO{4}\dO{5}\dO{6}
{1\over 6}
\Big\{ \xi(\gi\cd\gf) \Big[6 \xi(\gj\cd\gfi)\xi(\gk\cd\gs)+
                      9 \xi(\gj\cd\gk)\xi(\gfi\cd\gs)\Big]
+ \mbox{5 terms} \Big\} E_{\rm S}^\alpha(\gf,\gfi,\gs)
= 
\sum_{\alpha'} \lambda^{\alpha}_{\alpha'}
R_{\rm S}^{\alpha'}(\gi,\gj,\gk) ,
\label{eqq1}
\ee
expression which is symmetric in the arbitrary directions
$\gi,\gj,\gk$ as it should (note the presence of 
$R_{\rm S}^{\alpha'}$).

We aim at getting an explicit expression for $E_{\rm
S}^\alpha(\gf,\gfi,\gs)$. A glance at the previous equation shows that
we need to multiply both sides of it by the inverse of the correlation
function $\xi^{-1}$ defined in Appendix. Concretely,
we multiply Eq. (\ref{eqq1}) by
$\xigi{i}{i'}\xigi{j}{j'}\xigi{k}{k'}$
and integrate over directions $\gi , \gj , \gk$; we get
\bea
& 6 & E_{\rm S}^\alpha({\bf e}_{i'},{\bf e}_{j'},{\bf e}_{k'})
\nonumber\\
& + & 9 {(2\pi)^2\over 3}
\int\!\!\int\dO{5}\dO{6}
\xig{5}{6}
\Big[
\xigi{i'}{j'} E_{\rm S}^\alpha({\bf e}_{k'},\gfi,\gs)+
\xigi{j'}{k'} E_{\rm S}^\alpha({\bf e}_{i'},\gfi,\gs)+
\xigi{k'}{i'} E_{\rm S}^\alpha({\bf e}_{j'},\gfi,\gs)
\Big]
\nonumber\\
& = & 
{1\over 2}
\int\!\!\int\!\!\int\dO{i}\dO{j}\dO{k}
\xigi{i}{i'}\xigi{j}{j'}\xigi{k}{k'}
\sum_{\alpha'} \lambda^{\alpha}_{\alpha'}
R_{\rm S}^{\alpha'}(\gi,\gj,\gk) .
\label{eqq2}
\eea
As before, the $2\pi$ factors in the left hand side come from
operations like  $2\pi f(\gi)= \int\dO{k}
\delta(\gi\cd\gk - 1) f(\gk)$, for an arbitrary function $f(\gk)$.
However, we don't have an explicit expression for $E_{\rm S}^\alpha$ yet. 
We see that expressions of the type $\xi(\gfi\cd\gs) \xi^{-1}( {}_{-} \cd {}_{-} )
E_{\rm S}^\alpha( {}_{-} , \gfi , \gs )$ are the ones that prevent us from 
isolating the weight function.
To deal with this, it is convenient to construct the combination
$\int\!\!\int\dO{5}\dO{6} E_{\rm S}^\alpha( {}_{-} , {\bf e}_{5} , {\bf
e}_{6} ) \xig{5}{6}$.  To reach this goal, we multiply both sides of
Eq. (\ref{eqq2}) by $\xig{j'}{k'}$ and integrate over directions ${\bf
e}_{j'} {\rm and~} {\bf e}_{k'}$.  This operation produces a
divergence in the left hand side of this equation in a form of a Dirac
function `$\delta(0)$'. 
In the continuous case, all methods lead to this unavoidable problem
and although it has already appeared in the literature (\cite{Hea98}),
it has never been treated so far.
That this divergence is a mathematical artifact we can see from
the fact that, in practice, we never deal with an ideal experiment:
the problem is solved when we take into account the fact that
each different experimental setting is limited by a finite angular
resolution. This is usually quantified in terms of an $\ell$-dependent
window function ${\cal W}_\ell$ (the circularly symmetric pattern of
the observation beam in $\ell$-space), although more involved scanning
techniques are also employed (\cite{ws95,knox99}). 
Then, only a finite number of multipoles
will effectively contribute to the correlation function and, as a
result, the above mentioned divergence is regularized; indeed, we
have
\be
\xi(\gi\cd\gj) = \sum_\ell {2\ell+1\over 4\pi} {\cal C}_\ell 
{\cal W}^2_\ell P_\ell(\gi\cd\gj) ,
\ee
which, upon using the expression for $\xi^{-1}$ 
given in the Appendix, leads to the quantity
\be
{\cal A}_{i k}\equiv 
\int\dO{j} \xi(\gi\cd\gj) \xi^{-1}(\gj\cd\gk) 
= 2\pi \sum_{\ell m} {\cal W}^2_\ell \Ylmn(\gi)\Ylmn{}^*(\gk) .
\ee
In particular, the previous divergence `$\delta(0)$' now becomes 
\be
{\cal A} \equiv
{\cal A}_{i i} = 
\sum_{\ell} {2\ell+1\over 2} \, {\cal W}^2_\ell .
\label{AAA}
\ee 
This is a more realistic and finite object to work with in the case
where two directions on the microwave sky coincide for a given
experience.  Notice that for an ideal experimental setting in which
the window function ${\cal W}_\ell \to 1$, or equivalently the beam
$\sigma \to 0$ in the case of a Gaussian profile, ${\cal A}$ blows up.
Since the terms of the type $9 \xi \xi$ in Eq. (\ref{bougnoula}) 
are not present in the case of the two-point correlators, this problem 
did not appear there.
{}From now on, strictly speaking, all expressions should incorporate the
window function. However, in what follows and for computational
convenience, we will keep ${\cal A}_{i k} \approx \delta(\gi\cd\gk -
1)$ for $\gi \not= \gk$, a good approximation as we can see from
Fig. 2. 

\begin{figure}[t]
\centerline{\psfig{file=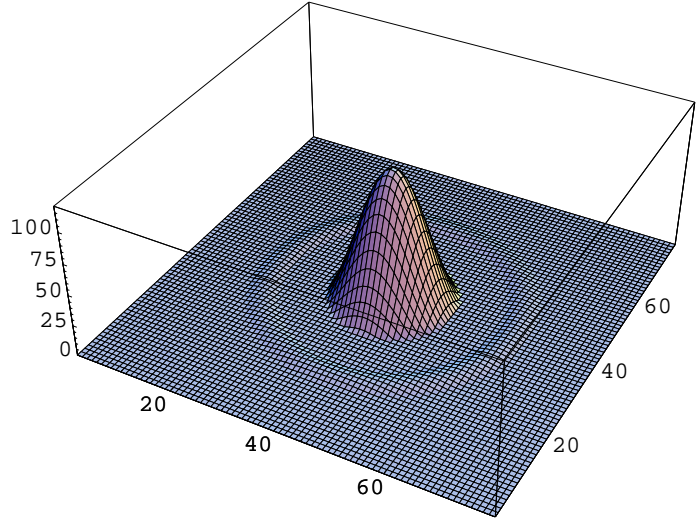,width=3.6in}}
\caption{Values of ${\cal A}_{i k}$ (as defined in the text) with
fixed direction $\gi$ at the center of the plot and direction 
$\gk$ scanning a square of side $90^\circ$. 
We show the result for the COBE-DMR window function.
The full width at half maximum of the plot is roughly $10^\circ$,
of the same order as the resolution of the COBE-DMR detector.
We can then expect the plotted `spike' to select just one experimental 
pixel on the sky map, making the relation 
${\cal A}_{i k} \approx \delta(\gi\cd\gk - 1)$ a good approximation.
Note that higher resolution experiments will yield a more
peaked curve, although one would expect the goodness of the approximation 
to remain roughly the same.}
\label{fig1}
\end{figure}

Endowed now with the above regularization method, we present
the term $\int\!\!\int\dO{5}\dO{6} E_{\rm S}^\alpha( {}_{-} , {\bf e}_{5} ,
{\bf e}_{6} ) \xig{5}{6}$ in the following form
\be
\int\!\!\int\dO{5}\dO{6}
E_{\rm S}^\alpha({\bf e}_{4},{\bf e}_{5},{\bf e}_{6})
\xig{5}{6} 
=
{{\cal D}\over 24 \pi^2}
\sum_{\alpha'} \lambda^{\alpha}_{\alpha'}
\int\!\!\int\!\!\int\dO{i'}\dO{j'}\dO{k'}
\xigi{4}{i'}\xigi{j'}{k'} 
R_{\rm S}^{\alpha'}({\bf e}_{i'},{\bf e}_{j'},{\bf e}_{k'})
\label{eqq3}
\ee
with ${\cal D} \equiv {24\pi^3} [6+9 {(2\pi)^2\over 3}(2\times 2\pi+
4\pi {\cal A})]^{-1}$ where, as expected, the factor ${\cal A}$
appears explicitly.  Now, we just replace the six terms with prefactor
$9 \xi \xi$ in the left hand side of Eq. (\ref{eqq1}) with
Eq. (\ref{eqq3}) resulting in
\bea
\label{eqq4}
\int\!\!\int\!\!\int
&  &\!\! 
\dO{4}\dO{5}\dO{6}
\xig{i}{4} 
\xig{j}{5} 
\xig{k}{6} 
E_{\rm S}^\alpha({\bf e}_{4},{\bf e}_{5},{\bf e}_{6})
\\
& = & 
{1\over 12}
\sum_{\alpha'} \lambda^{\alpha}_{\alpha'}
\Big\{
R_{\rm S}^{\alpha'}({\bf e}_{i},{\bf e}_{j},{\bf e}_{k})
\nonumber\\
& - &
{ {\cal D} \over 2\pi }
\int\!\!\int\dO{j'}\dO{k'}
\xigi{j'}{k'}
\Big[
R_{\rm S}^{\alpha'}({\bf e}_{i},{\bf e}_{j'},{\bf e}_{k'})
\xig{j}{k}+
R_{\rm S}^{\alpha'}({\bf e}_{j},{\bf e}_{j'},{\bf e}_{k'})
\xig{k}{i}+
R_{\rm S}^{\alpha'}({\bf e}_{k},{\bf e}_{j'},{\bf e}_{k'})
\xig{i}{j}
\Big]
\Big\} .
\nonumber
\eea
This contains just one appearance of $E^\alpha$; to get the weight
function explicitly, we only need to multiply the equation by
$\xigi{i}{1}\xigi{j}{2}\xigi{k}{3}$ and integrate over directions $\gi
, \gj , \gk$.  Finally, we get the expression for $E_{\rm S}^\alpha$
in terms of the Lagrange multipliers $\lambda^\alpha$
\bea
\label{eqq5}
&&
E_{\rm S}^\alpha({\bf e}_{1},{\bf e}_{2},{\bf e}_{3})
=  
{1\over 12 \times (2\pi)^3}
\sum_{\alpha'} \lambda^{\alpha}_{\alpha'}
\Big\{
\int\!\!\int\!\!\int\dO{i}\dO{j}\dO{k}
\xigi{i}{1}\xigi{j}{2}\xigi{k}{3}
R_{\rm S}^{\alpha'}({\bf e}_{i},{\bf e}_{j},{\bf e}_{k})
\\
& - & 
{\cal D}
\int\!\!\int\!\!\int\dO{i}\dO{j}\dO{k}
R_{\rm S}^{\alpha'}({\bf e}_{i},{\bf e}_{j},{\bf e}_{k})
\xigi{j}{k}
\Big[
\xigi{i}{1}\xigi{2}{3}+
\xigi{i}{2}\xigi{3}{1}+
\xigi{i}{3}\xigi{1}{2}
\Big]
\Big\}
\nonumber
\eea
Reached this point, we have an explicit expression for $E_{\rm
S}^\alpha$, but still dependent on the Lagrange multipliers.  This
equation is well defined as it contains the renormalized quantity
${\cal A}$. Within a particular experiment with a given resolution,
the value ${\cal A}$ takes depends on what one means by two coincident
directions. For example, for the COBE-DMR window-function
specification (a Gaussian beam with dispersion
$\sigma=3^\circ\llap.2$) this yields roughly ${\cal A} \approx 158.5$,
including the quadrupole.  It is not difficult to extend this to other
scanning techniques.  The previous equation is the analogue of our
Eq. (\ref{weight2H}) corresponding to the second moment and also to
Eq. (21) of (\cite{Hea98}).  In that article, a similar analysis is
done but for the discretized CMB sky.  Remark that no divergence
appears in his case.  Indeed, all relevant quantities are finite when
evaluated for two directions pointing towards the same pixel. Our
prefactor $\cal D$ corresponds to $3 / (2+3 N)$ in that paper, where
$N$ represents the number of pixels in the map.  Clearly, $N\to
\infty$ when the pixel size goes to zero, as well as $\cal A\to
\infty$ when the window function ${\cal W}_\ell \to 1$.  At this
intermediate step our corresponding expressions need not coincide
because both depend on the particular regularization scheme used (be
it discretization or usage of a window function).  Despite
appearances, we will show below that the final expression for the best
unbiased estimator does not depend on these schemes.  This cannot be
inferred from Eq. (\ref{eqq5}) because we still need to remove the
Lagrange multiplier. Unlike what was done in (\cite{Hea98}), we now
proceed further and express the weight function explicitly.
Hence, we multiply both sides of Eq. (\ref{eqq5})
by $R_{\rm R}^{\alpha''}({\bf e}_{1},{\bf e}_{2},{\bf e}_{3})$,
where $\alpha''\equiv
{\tiny\{\begin{array}{ccc}
         \! \ell_1'' \! & \! \ell_2'' \!  & \! \ell_3'' \\
         \! m_1''    \! & \! m_2''    \!  & \! m_3''    
                           \end{array}\}}$, 
then integrate over directions $\go$, $\gt$ and $\gth$ and, upon using 
the constraint equation (\ref{constraint}), we get
\bea
\label{bagarre}
 12  \delta_{\rm S}^{\alpha \alpha''} 
&=& 
 {1\over 6}
 {1\over 
  {\cal C}_{\ell_1''}{\cal C}_{\ell_2''}{\cal C}_{\ell_3''} }
\Big\{
\Big[
(-1)^{m_1''+ m_2''+ m_3''} 
\lamu{\ell_1''}{\ell_2''}{\ell_3''}{-m_1''}{-m_2''}{-m_3''}{\alpha}
+ \lamu{\ell_1''}{\ell_2''}{\ell_3''}{m_1''}{m_2''}{m_3''}{\alpha}
\Big]
+ \mbox{\rm 5 additional permutations in index $\alpha''$}
\Big\}
\nonumber 
\\
& - & 
{{\cal D} \over 3} 
\Biggl\{
\Biggl[
{\delta_{\ell_1'' \ell_2''}\delta_{m_1'' -m_2''} 
(-1)^{m_2''}
 \over 
 {\cal C}_{\ell_2''}{\cal C}_{\ell_3''}}
\sum_{\ell m}{(-1)^{m}\over {\cal C}_{\ell}}
\Big( 
(-1)^{m_3''}\lamu{\ell}{\ell}{\ell_3''}{m}{-m}{-m_3''}{\alpha}+
\lamu{\ell}{\ell}{\ell_3''}{-m}{m}{m_3''}{\alpha}+
(-1)^{m_3''}\lamu{\ell_3''}{\ell}{\ell}{-m_3''}{m}{-m}{\alpha}+
\lamu{\ell_3''}{\ell}{\ell}{m_3''}{-m}{m}{\alpha}
\nonumber
\\
& + &
(-1)^{m_3''}\lamu{\ell}{\ell_3''}{\ell}{m}{-m_3''}{-m}{\alpha}+
\lamu{\ell}{\ell_3''}{\ell}{-m}{m_3''}{m}{\alpha}  
\Big)\Biggr] 
+
{\tiny \Big[\begin{array}{c} 1\to 2 \\ 2\to 3 \\ 3\to 1 \end{array} \Big]}
+
{\tiny \Big[\begin{array}{c} 1\to 3 \\ 2\to 1 \\ 3\to 2 \end{array} \Big]}
\Biggr\} .
\eea
Eq. (\ref{bagarre}) is the final algebraic equation that the multipliers
must satisfy. For fixed $\alpha$, a natural way to 
proceed would be to get an explicit expression for 
$\lambda^\alpha_{\alpha'}$.
Another way to solve the problem goes on a line analogous to
the case of the two-point correlators: we just need to
identify the complicated combination of Lagrange multipliers in the 
right hand side of Eq. (\ref{bagarre}) with the one in the right hand side 
of Eq. (\ref{eqq4}) [or, equivalently, Eq. (\ref{eqq5})].
In order to do that, we now multiply both sides of Eq. (\ref{bagarre}) by
$R_{\rm R}^{\alpha''}({\bf e}_{i},{\bf e}_{j},{\bf e}_{k})$, perform the six 
sums over the indices in ${\alpha''}$ and we end up with
\bea
&&
R_{\rm S}^{\alpha}({\bf e}_{i},{\bf e}_{j},{\bf e}_{k})
{\cal C}_{\ell_1}{\cal C}_{\ell_2}{\cal C}_{\ell_3}
= 
\sum_{\alpha''}
R_{\rm R}^{\alpha''}({\bf e}_{i},{\bf e}_{j},{\bf e}_{k})
{\cal C}_{\ell_1''}{\cal C}_{\ell_2''}{\cal C}_{\ell_3''}
\delta_{\rm S}^{\alpha \alpha''}
\nonumber\\
& = &  
{1\over  12}
\sum_{\alpha'} \lambda^{\alpha}_{\alpha'}
R_{\rm S}^{\alpha'}({\bf e}_{i},{\bf e}_{j},{\bf e}_{k})
- 
{{\cal D}\over 36}
\sum_{L M}
\sum_{\ell m}
{(-1)^M\over {\cal C}_L }
\Big[
Y_{\ell}^{m}(\gi) \xig{j}{k}+
Y_{\ell}^{m}(\gj) \xig{k}{i}+
Y_{\ell}^{m}(\gk) \xig{i}{j}
\Big]
\nonumber\\
& \times &
\Big[
\lamu{\ell}{L}{L}{m}{M}{-M}{\alpha}+
(-1)^m \lamu{\ell}{L}{L}{-m}{-M}{M}{\alpha}+
\lamu{L}{\ell}{L}{-M}{m}{M}{\alpha}+
(-1)^m \lamu{L}{\ell}{L}{M}{-m}{-M}{\alpha}+
\lamu{L}{L}{\ell}{M}{-M}{m}{\alpha}+
(-1)^m \lamu{L}{L}{\ell}{-M}{M}{-m}{\alpha}
\Big] .
\label{bagarrenew}
\eea
This is the equivalent of Eq. (\ref{LH2}). The aim now is to show that the 
combination of Lagrange multipliers in the right hand side of the previous 
equation is precisely the one which appears in the right hand side of  
Eq. (\ref{eqq4}).
In the latter, let us express both $R_{\rm S}^{\alpha'}$ 
and $\xi^{-1}$ in the second term in the right hand side in terms of 
spherical harmonics. After some algebra we get
\bea
\int\!\!\int\!\!\int
&  &\!\! 
\dO{4}\dO{5}\dO{6}
\xig{i}{4} 
\xig{j}{5} 
\xig{k}{6} 
E_{\rm S}^\alpha({\bf e}_{4},{\bf e}_{5},{\bf e}_{6})
\nonumber\\
& = & 
{1\over 12}
\sum_{\alpha'} \lambda^{\alpha}_{\alpha'}
R_{\rm S}^{\alpha'}({\bf e}_{i},{\bf e}_{j},{\bf e}_{k})
-  
{{\cal D}\over 36}
\sum_{L M}
\sum_{\ell m}
{(-1)^M\over  {\cal C}_L }
\Big[
Y_{\ell}^{m}(\gi) \xig{j}{k}+
Y_{\ell}^{m}(\gj) \xig{k}{i}+
Y_{\ell}^{m}(\gk) \xig{i}{j}
\Big]
\nonumber\\
& \times &
\Big[
\lamu{\ell}{L}{L}{m}{M}{-M}{\alpha}+
(-1)^m \lamu{\ell}{L}{L}{-m}{-M}{M}{\alpha}+
\lamu{L}{\ell}{L}{-M}{m}{M}{\alpha}+
(-1)^m \lamu{L}{\ell}{L}{M}{-m}{-M}{\alpha}+
\lamu{L}{L}{\ell}{M}{-M}{m}{\alpha}+
(-1)^m \lamu{L}{L}{\ell}{-M}{M}{-m}{\alpha}
\Big] ,
\label{eqq4new}
\eea
which, as advertised, yields the same combination of Lagrange
multipliers of Eq. (\ref{bagarrenew}).  Then, putting the last two
equations together, multiplying by $\xi^{-1}$ and integrating three
times, we finally get the weight function associated to the best
unbiased estimator
\be
E_{\rm S, Best}^\alpha(\gi,\gj,\gk) =  R_{\rm S}^{\alpha}(\gi,\gj,\gk) ,
\ee
which implies that the best unbiased estimator itself is given by
\be
{\cal E}_{\rm Best}({\cal B}^{\alpha })= 
\frac{1}{2}
\Big(
\alm{1}\alm{2}\alm{3} +
a_{\ell_1}^{m_1 *}
a_{\ell_2}^{m_2 *} 
a_{\ell_3}^{m_3 *} \Big) 
.
\label{newthird}
\ee
This is the final answer and it is a new result.  Let us make a few
remarks.  Firstly, this does not depend on ${\cal A}$ which shows that
Eq. (\ref{newthird}) is independent of the regularization scheme used.
Secondly, Luo (1994) used the following complex unbiased estimator: 
${\cal E}({\cal B}^{\alpha })=\alm{1}\alm{2}\alm{3}$, 
although he did not claim it to be the best one. 
Thirdly, as for the two-point correlators, one cannot use Eq. (\ref{newthird})
in order to infer ${\cal E}_{\rm Best}({\cal C}_{\ell_1 \ell_2 \ell_2})$.
Indeed, promoting the equation 
\be
\bigl\langle a_{\ell_1 }^{m_1} a_{\ell_2}^{ m_2} a_{\ell_3}^{ m_3}
\bigr\rangle =
\wjm
{\cal C}_{\ell_1 \ell_2 \ell_3 } ,
\ee
valid for the mean values of the estimators, to the following equation 
\begin{equation}
\label{EHbestwrong2}
{\cal E}_{\rm Best}({\cal B}^{\alpha})
= 
\frac{1}{2}
\Big(
\alm{1}\alm{2}\alm{3} +
a_{\ell_1}^{m_1 *}
a_{\ell_2}^{m_2 *} 
a_{\ell_3}^{m_3 *} \Big) 
= {\cal E}_{\rm Best}({\cal C}_{\ell_1 \ell_2 \ell_3}) \wjm
\quad \mbox{(false)} ,
\end{equation}
valid for the estimators themselves, is an unjustified step. 
If, nevertheless, we used this false relation we would get
\begin{equation}
\label{EHbestwrong22}
{\cal E}_{\rm Best}({\cal B}^{\alpha})
= 
\wjm
\sum _{m_1' m_2' m_3'}
\wjma{\ell_1 }{\ell_2 }{\ell_3 }{m_1'}{m_2'}{m_3'}
a_{\ell_1 }^{m_1'}a_{\ell_2 }^{m_2'}a_{\ell_3 }^{m_3'}
\quad \mbox{(false)} ,
\end{equation}
which cannot be cast into 
$\frac{1}{2}
\Big(
\alm{1}\alm{2}\alm{3} +
a_{\ell_1}^{m_1 *}
a_{\ell_2}^{m_2 *} 
a_{\ell_3}^{m_3 *} \Big)$.
So, like for the two-point correlators, we see that one cannot 
infer the ${\cal E}_{\rm Best}({\cal B}^{\alpha})$ from 
${\cal E}_{\rm Best}({\cal C}_{\ell_1 \ell_2 \ell_3})$ and vice versa.

Endowed now with the best unbiased estimator, one can compute its 
variance squared, the `third-moment cosmic variance', which reads
\bea
\sigma^2_{{\cal E}_{\rm Best}({\cal B}^{\alpha })}
& = & 
\frac{1}{2}
\Big\{
{\cal C}_{\ell_1}{\cal C}_{\ell_2}{\cal C}_{\ell_3}
\Big[1+ 
\delta_{m_1   0}
\delta_{m_2   0}
\delta_{m_3   0}
\Big]
+
{\cal C}_{\ell_1}^3\delta _{\ell_1\ell_2\ell_3}
\Big[
8 
\delta_{m_1   0}
\delta_{m_2   0}
\delta_{m_3   0}
+
2
(\delta _{m_1m_3}+\delta _{m_1 -m_3})
(\delta _{m_1m_2}+\delta _{m_1 -m_2})
\Big]
\nonumber \\
&+&
{\cal C}_{\ell_1}{\cal C}_{\ell_2}^2
\delta _{\ell_2\ell_3}
\Big[
2 \delta_{m_1   0} \delta_{m_2 -m_3}
+ \delta_{m_2m_3} + \delta_{m_2 -m_3}
\Big]
+
{\cal C}_{\ell_2}{\cal C}_{\ell_3}^2
\delta _{\ell_3\ell_1}
\Big[
2 \delta_{m_2   0} \delta_{m_3 -m_1}
+ \delta_{m_3 m_1} + \delta_{m_3 -m_1}
\Big]
\nonumber \\
& + &
{\cal C}_{\ell_3}{\cal C}_{\ell_1}^2
\delta _{\ell_1\ell_2}
\Big[
2 \delta_{m_3   0} \delta_{m_1 -m_2}
+ \delta_{m_1 m_2} + \delta_{m_1 -m_2}
\Big]
\Big\} .
\eea 
For example, from this we can now compute the cosmic variance
for the third-moment estimator ${\cal E}_{\rm Best}({\cal B}^\beta)$
with 
$\beta\equiv
{\tiny\{\begin{array}{ccc}
         \! \ell    \! & \! \ell    \!  & \! \ell   \\
         \!  m      \! & \!  m      \!  & \! -2m
                           \end{array}\}}$ 
where $\ell = \mbox{even}$ and $m\neq 0$. 
This particular case is often treated in the literature, 
see e.g. (Luo 1994, Heavens 1998).
We find 
$\sigma^2_{{\cal E}_{\rm Best}({\cal B}^{\beta})} = 
{\cal C}^3_{\ell}$ whereas the variance of the 
estimator used in (Luo 1994) yields 
$\sigma^2_{{\cal E}({\cal B}^{\beta})} = 2 \, {\cal C}^3_{\ell}$.
Another example comes from taking
$\gamma\equiv
{\tiny\{\begin{array}{ccc}
         \! \ell    \! & \! \ell    \!  & \! \ell   \\
         \!  m_1    \! & \!  m_2    \!  & \! m_3
                           \end{array}\}}$ 
where $|m_i|\neq |m_j|$ for any $i, j$. With this choice
we get 
$\sigma^2_{{\cal E}_{\rm Best}({\cal B}^{\gamma})} 
= {\cal C}^3_{\ell} / 2$ whereas 
Luo (1994) obtains 
$\sigma^2_{{\cal E}({\cal B}^{\gamma})} =  {\cal C}^3_{\ell}$.
Note that in both examples the results differ by a 1/2 factor. 
This can be traced back to the form of the best estimator in 
Eq. (\ref{newthird}).
The variance computed in (Luo 1994) is consistent with his 
choice of the estimator. However, unlike what is stated in that paper,
this variance does not deserve the name `cosmic' because, as we saw 
above, this estimator is not the best one. As expected, the variance 
of the best estimator is smaller than the variance computed in his 
article.
Since we have now the correct expression for the cosmic variance, 
its numerical value should be re-estimated. 
To be specific, let us take 
$\beta\equiv
{\tiny\{\begin{array}{ccc}
         \!  2    \! & \! 2    \!  & \! 2   \\
         \!  1      \! & \!  1      \!  & \! -2
                           \end{array}\}}$.
The cosmic variance is then 
$\sigma_{{\cal E}_{\rm Best}({\cal B}^{\beta})} =
{\cal C}^{3/2}_{2} = (4\pi / 5)^{3/2} Q_{\rm rms-PS}^3 / T_0^3
\approx 1.3 \times 10^{-15}$ where we used 
$T_0 = 2.7$ K and $Q_{\rm rms-PS} = 18.7 \mu$K (\cite{Bunn97}).
Although this figure is close to Luo's result 
($1.4 \times 10^{-15}$) cited after equation (32) in 
(Heavens 1998), this does not imply that the two variances are 
not different by a factor 1/2, as we have just seen.
This might probably be due to a difference in the quadrupole 
normalizations.

\section{Conclusions}

Optimized analyses of CMB datasets involve the use of appropriate
methods in order to reduce the various uncertainties. In particular,
the theoretical error bars due to the cosmic variance can be minimized
by working with the method of the best unbiased estimators.  In this
article, we have applied this technique for the study of CMB
non-Gaussian features. These are often characterized by means of the
third moment for the $a^m_\ell$'s or by the angular bispectrum ${\cal
C}_{\ell_1\ell_2\ell_3}$. We have found the best unbiased estimators
in both cases. These are the quantities that should be used in future
data analyses and would be important for upcoming megapixel experiments 
(\cite{t97,b99}) like MAP\footnote{{\tt http://map.gsfc.nasa.gov/ }} 
and Planck Surveyor\footnote{{\tt http://astro.estec.esa.nl/Planck/ }}.  
In addition to this, we have displayed both 
the angular bispectrum and the third moment cosmic variances, 
the smallest possible uncertainties attached to the bispectrum and
the third moment, which would be present in any ideal experiment when 
all other sources of noise have been removed.

\section*{Acknowledgments}

A.G is member of CONICET, Argentina.

\section{Appendix: The inverse two-point correlation function}

In this Appendix, we derive the exact expression of the inverse of the 
two-point correlation function $\xi ^{-1}$, defined according to
\begin{equation}
\label{definv}
\int {\rm d}\Omega _j \xi ({\bf
e}_i\!\cdot\!{\bf e}_j)\xi ^{-1}({\bf e}_j\!\cdot\!{\bf e}_k)\equiv \delta
({\bf e}_i \cd {\bf e}_k-1) .  
\end{equation}
As usual, one expands the two-point correlation function on the basis of
the Legendre polynomials 
\be
\xi(\gi\cd\gj) = \sum_\ell {2\ell+1\over 4\pi}{\cal C}_\ell
P_\ell(\gi\cd\gj) .
\label{xiapp}
\ee
In general, $\xi ^{-1}$ can be expanded on the spherical harmonics 
basis as follows
\begin{equation}
\label{expY}
\xi ^{-1}({\bf e}_i\!\cdot\!{\bf e}_j)=\sum _{\ell m}\sum _{\ell ' m'}
b_{\ell \ell' m m'}Y_{\ell }^{m}({\bf e}_i)Y_{\ell '}^{m'*}({\bf e}_j).
\end{equation}
Our aim now is to determine the coefficients $b_{\ell \ell' m m'}$. 
Using Eqns. (\ref{xiapp}) and (\ref{expY}), 
the completeness relation for the Legendre polynomials
together with the addition theorem of spherical harmonics 
in Eq. (\ref{definv}), one gets 
\begin{equation}
\label{intinv}
\sum _{\ell m}\sum _{\ell ' m'}{\cal C}_{\ell }
b_{\ell \ell' m m'}Y_{\ell }^{m}({\bf e}_i)Y_{\ell '}^{m'*}({\bf e}_k)=
\sum _{\ell ''}(\ell ''+\frac{1}{2})P_{\ell ''}
({\bf e}_i\cdot {\bf e}_k) .  
\end{equation}
The coefficients $b_{\ell \ell' m m'}$ are easy to read off and one obtains 
\begin{equation}
\label{coeffb}
b_{\ell \ell' m m'}=\frac{2\pi }{{\cal C}_{\ell }}\delta _{\ell \ell'}\delta _{mm'},
\end{equation}
from which one deduces
\begin{equation}
\label{finalinv}
\xi ^{-1}({\bf e}_i\!\cdot\!{\bf e}_j)=\sum _{\ell }
\frac{\ell +1/2}{{\cal C}_{\ell }}P_{\ell }( {\bf e}_i\cdot {\bf e}_j) .
\end{equation}
This is the expression used in the main text.

\end{document}